\documentclass[1p,longtitle,sort&compress]{elsarticle}

\journal{Physica A}

\usepackage{amssymb,amsmath}
\usepackage{graphicx}% Include figure files
\usepackage{subfigure}%
\usepackage{bm}% bold math
\usepackage{cases}
\usepackage{verbatim} % Include multi-linear comments
\usepackage{color}

\newcommand\beq{\begin{equation}}
\newcommand\eeq{\end{equation}}
\newcommand\beqa{\begin{eqnarray}}
\newcommand\eeqa{\end{eqnarray}}
\newcommand{\nn}{\nonumber\\}
\def\bal#1\eal{\begin{align}#1\end{align}}

\newcommand{\PP}{\mathcal{P}}
\newcommand{\FP}{\mathcal{F}}
\newcommand{\pad}{\text{Pad\'e}}

\begin{document}
\begin{frontmatter}

\title{Random Walks on Lattices. Influence of Competing Reaction Centers on Diffusion-Controlled Processes}

\author{Enrique Abad\corref{mycorrespondingauthor}\fnref{myfootnote}}
\ead{eabad@unex.es}
\address{
Departamento de F\'isica Aplicada (Centro Universitario de M\'erida) and Instituto de Computaci\'on Cient\'ifica Avanzada (ICCAEx), Universidad de Extremadura, E-06800 M\'erida, Spain}
\fntext[myfootnote]{Corresponding author.}
\author{Tim Abil}
\ead{tim.abilzade@gmail.com}
\address{College of Computing and Digital Media, DePaul University, 243 South Wabash, Chicago, Illinois 60604-2301, USA}
\author{Andr\'es Santos}
\ead{andres@unex.es}
%\ead[url]{http://www.unex.es/fisteor/andres/}
\address{Departamento de F\'isica and Instituto de Computaci\'on Cient\'ifica Avanzada (ICCAEx),  Universidad de Extremadura, E-06006 Badajoz, Spain}
\author{John J. Kozak}
\ead{kozak@depaul.edu}
\address{Department of Chemistry, DePaul University, 243 South Wabash, Chicago, Illinois 60604-2301, USA}

\date{\today}

\begin{abstract}
We study diffusion-reaction processes on periodic simple cubic (sc) lattices and square planar lattices. We consider a single diffusing reactant undergoing an irreversible reaction upon first encounter with a static co-reactant placed at a given site (``one-walker problem"). We also allow for a competing reaction, namely, instantaneous trapping of the diffusing reactant at any other site (with probability $s$) before interacting with the static co-reactant. We use a generating function approach and Markov theory, as well as MC simulations, to determine the mean walklength $\langle n \rangle$ of the diffusing reactant before either of the two competing reactions takes place. To investigate the dependence of $\langle n \rangle$ on lattice size we compute the first, finite size corrections to the Green function of the sc lattice, correcting results reported in the literature. Using these results, space exploration properties and first-passage properties (e.g. walklength statistics, mean number of distinct sites visited, statistics of return to the origin, etc.) of both conventional (immortal) walks and mortal walks can be determined. In this context, we develop a novel approach based on a two-point Pad\'e approximant for the Green function. Finally, we study by means of MC simulations the more complex case where both reactant and co-reactant undergo synchronous nearest-neighbor displacements (``two-walker problem"). Here, we assume that reactant and co-reactant can individually be trapped with probability $s$ at any lattice site, or can undergo an irreversible reaction on first encounter at any site. When $s=0$ we find that, both for the one-walker and the two-walker problem, for lattices with (approximately) the same number of sites, the mean walklength is smaller (and hence the reaction efficiency greater) in $d=3$ than in $d=2$. Increasing $s$ tends to reduce differences in system dimensionality, and distinctions between the one-walker problem and the two-walker problem. Our model provides a good starting point to develop studies on the efficiency of apparently diverse diffusion-reaction processes, such as diffusion on a partially poisoned catalytic substrate or photosynthetic trapping of excitations.
\end{abstract}

\end{frontmatter}

%\linenumbers

\section{Introduction}
\label{sec1}
Lattice models provide an important paradigm for the study of phenomena involving diffusion-controlled reactions. One of the simplest models consists of a single diffusing reactant undergoing a nearest-neighbor random walk on a $d=2$ Euclidean lattice.  One site of the lattice is regarded as a (static) reaction center. As reviewed by Weiss \cite{W94}, an extensive literature deals for the case where the diffusing reactant is assumed to undergo an irreversible trapping reaction on first encounter at the active site (static trap). This setting corresponds to what we term ``one-walker problem", since only one of the two reaction partners is mobile. Beyond its numerous possible applications, the one-walker problem has historically played a central role in the theory of random walks, notably in the development of powerful methods relying on generating functions. Of fundamental importance in this context is the computation of lattice Green functions, whence many characteristic space exploration properties of walks in discrete and continuous time can be derived \cite{H95, KMS73}.

 In real systems, one usually does not have a single active site, but rather a variety of active sites which compete with one another. As a result of this competition, the time scale of the reaction is significantly reduced. In general, the trapping probability $s_i$ of each of these reaction centers is different. See, for example, the seminal paper by den Hollander and Kasteleyn \cite{HK82}.

The case where a perfect trap at one lattice site ($s=1$) coexists with a non-zero probability of absorption ($0<s<1$) at all other lattice sites (also termed background trapping in what follows) lends itself particularly well to the analysis of finite size effects \cite{AK15}. Although the above model is of interest for a variety of different situations, a possible interpretation of non-zero background trapping is site deactivation due to partial poisoning of a catalyst.  In the above  simplified setting, there is a single fully active site or deep trap, whereas all other sites are only partially active. As argued in Ref.\ \cite{AK15}, differences in the catalytic activities of individual sites may be ascribed to differences in the affinity with a chemisorbed poisoning species, resulting in selective poisoning \cite{G11}. Despite its simplicity, the above model, with a single site immune to poisoning and all other sites partially poisoned, efficiently illustrates the onset of a nonlinear decrease in the catalytic efficiency as a function of the concentration of the poisoning agent \cite{AK15}. As already mentioned, this setting also provides a suitable starting point to describe more realistic situations by introducing additional sites immune to poisoning \cite{PK85, SMK85}.

In the present work we study the above model in Euclidean dimension $d=3$ and compare our results with previous results for $d=2$ \cite{AK15}. In this context, we also consider the ``two-walker'' problem; that is, instead of assuming that the reaction center is static, we allow for the possibility that it is a moving target. In the language of chemical kinetics, we study the effect of competing reaction centers on diffusion controlled first- and second-order chemical processes occurring on surfaces or in three dimensions. Both the cases $s=0$ and $s>0$ are studied.   We report Monte Carlo (MC) simulations to corroborate and extend results obtained in our analytic studies.

It is worth noting that the computation of the walklength of a walker subject to background trapping is a particular example of a broad class of first-passage problems for so-called mortal walkers, i.e., walkers that have a finite probability of dying as they move. Recently, there has been a surge of interest in such problems \cite{YAL13, AYL13, YAL14, BR14, CAMBL15, MR15, RB15, OMR16, GR17, BRB17, B17, ER17, CBM17} motivated by a large number of applications, including photosynthetic trapping \cite{AK15}, oocyte fertilization \cite{BR14}, game theory \cite{AK15}, radioactive decay \cite{AK15, GR17}, target deactivation problems \cite{CAMBL15} or foraging behavior \cite{CBM17}.

While the physics of mortal random walkers may be very different from that of standard walkers, generating function approaches \emph{\`a la} Montroll--Weiss \cite{W64,MW65,W69,W94} still apply and can be used to obtain space exploration properties provided that the lattice remains translationally invariant.

The extension of the generating function approach to deal with ``mortal walkers'' \cite{YAL13, AYL13, YAL14, BR14, CAMBL15, MR15, RB15, OMR16, GR17, BRB17, B17, CBM17, ER17} has not only expanded the range of physical problems that can be explored using a lattice-based theory, but has
also motivated the development of novel approximations for lattice Green functions, as will be
illustrated later on in this study.  At present, however, an approach based on lattice Green functions can be implemented only for translationally-invariant lattices. To study diffusion-reaction processes on lattices with spatial (or other) defects, fractal lattices, or disordered lattices, one must rely on the classical theory of finite Markov processes and/or numerical experiments (MC or molecular dynamics simulations).  A distinguishing feature of this contribution is that we
implement all three strategies, viz.\ generating functions, Markov theory, and MC simulation,
to study the influence of competing reaction centers on the efficiency of diffusion-controlled reactions.

Finally, we note that in the present paper we consider the kinetics of a pair of coreactants both in the limit of large and small system sizes. The larger the system, the better Smoluchowski-like approaches based on continuum diffusion are expected to apply. However, for small systems (the $5\times 5$ lattice considered in Sec.\ \ref{sec2}, say), effects due to the discrete nature of the reactants as well as crowding effects arising from the finite system size become apparent, and significant departures from continuum approximations may appear.

The remainder of this paper is organized as follows. In Sec.\ \ref{sec2}, we first present the Markov method and show how to use it to compute the walklength of a mortal walker (representing a diffusing reactant) in the presence of a deep trap (representing a static reaction center) for the particular case of a periodic $5\times 5$ lattice. In Sec.\ \ref{sec3}, we introduce the generating function formalism for a $d$-dimensional hyperlattice, we give a general formula for the walklength in terms of the nonsingular part of the lattice Green function, and we illustrate the method by rederiving the result obtained in Sec.\ \ref{sec2} for the $5\times 5$ lattice.
Section \ref{sec4} is devoted to the computation of finite size corrections to the Green function of the periodic sc lattice; these results are the starting point to compute various approximations for the mean walklength of the mortal walker. These approximations are subsequently compared with exact results, which are in turn validated via MC simulation. In Sec.\ \ref{sec5}, we discuss the role of the lattice dimensionality and of the probability $s$ in the one-walker problem and in the two-walker problem. In this section, we also discuss the possible relevance of our results for real systems. Finally, we summarize the main conclusions and outline possible pathways for future research in Sec.\ \ref{sec6}. Technical details of the most relevant calculations are given in the appendices.

\section{Markovian approach}
\label{sec2}

Sixty years ago,  Montroll and Shuler published a review \cite{MS57}  describing a comprehensive  approach for studying theoretically chemical and physical  transformations.  Consider a reaction space of given size and shape,  characterized by $N$ discrete lattice points having local connectivity $v$ and embedded in a Euclidean space of dimension $d$.  The probability distribution function governing the fate of a diffusing particle satisfies a stochastic master equation whose kernel is related to the fundamental matrix of the theory of finite Markov processes.  In the limit of (sufficiently) large $N$, the first moment of the probability distribution function, which is just the mean walklength $\langle n \rangle$ of the Markovian theory, is related to the smallest eigenvalue of the stochastic master equation, and from thence to a chemical rate constant;  see the later review \cite{K00} and references cited therein.

The methods described in Ref.\  \cite{KS60} allow one to calculate quantities from random walk trajectories from any site $l$ to another site $j$ of a host lattice, and in particular, the total number of steps taken by a random walker before reaching site $j$, the average of this quantity over all sites $l\neq j$ (overall mean walklength), as well as any desired higher-order moment of the probability distribution function for the latter quantity.

In the present work, we are interested in studying the change in the efficiency of a trapping reaction when a diffusing reactant can be trapped before encountering a static co-reactant. The first application of the Markovian approach described above to this problem was presented in Refs.\ \cite{WK81,WK82}.  Exact numerical results were reported for a random walker on a $d$-dimensional Euclidean lattice of $N$ sites, in which a deep trap is placed at a given site and imperfect traps with trapping probability $s<1$ are placed at each of the remaining $N-1$ sites.

In Refs.\ \cite{WK81,WK82}, the above program was illustrated for a random walker on a $5\times5$ square-planar Euclidean lattice. Each site can be assigned a symmetry class depending on its relative position with respect to the deep centrosymmetric trap $T$. In the present case, apart from the centrosymmetric site $T$, if the boundary conditions are homogeneous there are five distinct symmetry classes, illustrated by the following scheme:
\beq
\label{sch}
\begin{array}{|c|c|c|c|c|}
\hline
5 & 4 & 3 & 4 & 5 \\ \hline
4 & 2 & 1 & 2 & 4 \\ \hline
3 & 1 & T & 1 & 3 \\ \hline
4 & 2 & 1 & 2 & 4 \\ \hline
5 & 4 & 3 & 4 & 5 \\ \hline
\end{array}
\eeq

In order to proceed further, one has to specify the boundary conditions. In Ref.\ \cite{AK15}, periodic boundary conditions were chosen, implying that the lattice becomes translationally invariant. Here, we stick to this choice, since we wish to keep our setting as simple as possible by assessing the influence of finite size effects only (for a discussion of boundary effects, see Ref.\ \cite{K00}). For each symmetry-distinct initial condition $i=1,\ldots,5$, let us denote by $\langle n \rangle_i$ the number of times a random walker performs site-to-site transitions before being trapped at site $T$ (site specific mean walklength). Let us further define $\langle n \rangle_T$ as the mean walklength when the walker starts from the deep trap. These quantities are related to one another via the following linear system:
\begin{subequations}
\label{2.1}
\beq
\langle n \rangle_1  = s + (1-s)\left(\frac{1}{4}\langle n \rangle_T + \frac{1}{2}\langle n \rangle_2 + \frac{1}{4}\langle n \rangle_3+ 1\right),
\eeq
\beq
\langle n \rangle_2  =  s + (1-s)\left(\frac{1}{2}\langle n \rangle_1 + \frac{1}{2}\langle n \rangle_4 + 1\right),
\eeq
\beq
\langle n \rangle_3 =  s + (1-s)\left(\frac{1}{4}\langle n \rangle_1 + \frac{1}{4}\langle n \rangle_3 + \frac{1}{2}\langle n \rangle_4 + 1\right),
\eeq
\beq
\langle n \rangle_4  =  s + (1-s)\left(\frac{1}{4}\langle n \rangle_2 + \frac{1}{4}\langle n \rangle_3 + \frac{1}{4}\langle n \rangle_4 + \frac{1}{4}\langle n \rangle_5 + 1\right),
\eeq
\beq
\langle n \rangle_5  =  s  + (1-s)\left(\frac{1}{2}\langle n \rangle_4 + \frac{1}{2}\langle n \rangle_5 + 1\right).
\eeq
\end{subequations}
 These equations reflect the fact that a particle starting at a given site undergoes one of two mutually exclusive events: either it dies upon taking the first step (in which case the walklength is equal to one) or else it continues its walk from another site which does not in general belong to the same symmetry class as the previous site. The first event is weighted with probability $s$ and is represented by the first term on the right hand side of the above equations. The second event (described by the last term on the right hand side) is weighted with the complementary probability $1-s$. Due to the absence of memory effects (Markovian hypothesis), this event effectively results in the random walk starting anew from any of the possible arrival sites (except for the fact that the walklength must be increased by one unit). In this term, the walklengths referring to each of these sites are weighted with the corresponding transition probabilities, which are directly obtained by inspecting the scheme \eqref{sch} [for a general formulation beyond the $5\times5$ case see Refs.\ \cite{WK81,WK82, AK15}].

 Unless otherwise specified, in what follows we shall assume that a walker initially placed at the deep trap $T$ is immediately absorbed. This allows one to solve for the $\langle n \rangle_i$'s by setting $\langle n \rangle_T=0$ in Eqs.\ \eqref{2.1}. The resulting expressions for the $\langle n \rangle_i$'s are given in \ref{app1}.  From these expressions, the overall mean walklength $\langle n \rangle$  before the walker is trapped can be calculated. The quantity $\langle n \rangle$ is defined as the average over all the possible starting sites (excluding site $T$) under the assumption that all of them have the same statistical weight. In general, the number of sites in each symmetry class is different and this needs to be taken into account when computing $\langle n \rangle$ \cite{WK81}. In the present case of a $5\times5$ lattice one eventually obtains [see \eqref{sch}]
\bal
\label{5b5wl}
\langle n \rangle=&\frac{\langle n \rangle_1+\langle n \rangle_2+\langle n \rangle_3+2\langle n \rangle_4+\langle n \rangle_5}{6}\nn
=&\frac{1}{3}\frac{475+500s-150s^2-60s^3+3s^4}
{5+161s+158s^2-50s^3-19s^4+s^5}.
\eal
In  Sec.\ \ref{sec3}, we will show how this result can be recovered by means of a generating function approach, which is especially well suited to assess finite size effects.

\section{Generating function approach}
\label{sec3}

The use of generating functions to study random walks on $d$-dimensional Euclidean lattices with a single deep trap was developed by Montroll and Weiss \cite{W64,MW65,W69}. More recently, a generating function approach for studying the consequences of introducing partial traps at the remaining $N-1$ sites was introduced to study the case of periodic lattices in $d=1$ and $d=2$ \cite{YAL13,YAL14}.  In this section, we recall some of the main results and give a general formula for the walklength in terms of the Green function of a finite $d$-dimensional hypercubic lattice.

Consider a P\'olya walk on a $d$-dimensional, translationally invariant lattice with $N=L^d$ sites and a deep trap placed at a given site (which we term the origin ``$\overrightarrow{0}$") and $N-1$ partially absorbing sites. The walker dies with probability one if it steps on the origin; when the walker steps on any other site, it either dies with probability $s$ or else it performs a jump to a nearest-neighbor site with complementary probability $1-s$.

We denote by  $\left\langle n\right\rangle _{\vec{\ell }} $ the trajectory-averaged walklength of a walker starting at a given lattice site $\vec{\ell }=(\ell _{1} ,\ell _{2} ,\ldots ,\ell _{d} )\ne \overrightarrow{0}$.  Let us further introduce the initial-condition averaged walklength (overall mean walklength)
\beq
\left\langle n\right\rangle =\frac{1}{N-1} \sum _{\vec{\ell }\ne \overrightarrow{0}}\left\langle n\right\rangle _{\vec{\ell }}
\eeq
to characterize the (coarse-grained) mean lifetime of this so-called mortal walk. In Ref.\ \cite{AK15}, the following general result was derived, viz.,
\begin{equation}
\label{genres}
\left\langle n\right\rangle =\frac{1-\Gamma (\overrightarrow{0},1-s)}{s}.
\end{equation}
(For $s=0$, the right-hand side should be interpreted as the limit $s\to 0$). The auxiliary quantity  $\Gamma (\overrightarrow{0},z)$  is defined as follows:
\beq
\label{auxeq}
\Gamma (\overrightarrow{0},z)\equiv \frac{1}{N-1} \left[\frac{1}{(1-z)\PP(\overrightarrow{0},z)} -1\right].
\eeq
Here,  $\PP(\overrightarrow{0},z)$ stands for the lattice Green function
\beq
\PP(\vec{0},z)=\sum _{n=0}^{\infty}\PP_{n} (\vec{0}) z^{n},
\eeq
defined as the generating function for the probability that a conventional (immortal) walk starting at the origin (lattice site where the deep trap is located) returns to it [$\PP_{n} (\vec{0})$  is the probability that the walker is found at the origin after exactly $n$ time steps].  In what follows, we shall use for convenience the short-hand notation $P_d(z)$ in place of $\PP(\vec{0},z)$. The explicit reference to the spatial dimension in our notation has some advantages that will become evident later on.

Note that the validity of Eq.\ (\ref{genres}) is not restricted to a P\'olya walk, since it also applies to non-nearest neighbor walks as long as the lattice remains translationally invariant. It is worth noting that the (initial-condition-averaged) probability $p_T$ for the walker to eventually be absorbed at the deep trap rather than at any other site is simply given by the expression $p_T=1-s \langle n \rangle $ (see Ref.\ \cite{AK15}). This probability is simply $p_T=\sum_{n=0}^\infty p_{T,n}$, where $p_{T,n}$ is the probability of absorption after \emph{exactly} $n$ time steps (this quantity is the discrete analogue of the particle flux towards the trap in the framework of a continuum approximation). As shown in \ref{app2}, one obtains
\beq
\label{EqpTn}
p_{T, n}=\frac{(1-s)^n}{n! (N-1)}\frac{\partial^n}{\partial z^n} \left. \left\{ (1-z) P_d(z)\right\}^{-1}\right|_{z=0}
\eeq
The generating function $P_d(z)$ can be separated into a part that diverges as $z\to 1^-$ (singular part) and a nonsingular part $Q_d(z)$ which remains finite in this limit, i.e.,
\beq
\label{3.5}
Q_d(z)=P_d(z)-\frac{1}{N(1-z)}.
\eeq
From Eq.\ \eqref{genres}, it is then easy to show that the mean walklength $\langle n \rangle$ is given by
\beq
\label{QNwl}
\frac{\langle n \rangle}{N}=\frac{N}{N-1}\frac{Q_d(1-s)}{1+Ns Q_d(1-s)},
\eeq
whence the small $s$-expansion
\begin{equation}
\label{expQs}
\frac{\langle n \rangle}{N}=\frac{N}{N-1}Q_d(1) -\frac{N}{N-1}\left[Q_d^2(1)+\frac{Q_d'(1)}{N}\right]Ns+\mathcal{O}(s^2)
\end{equation}
follows. The above expansion turns out to be useful for the study of the small $s$ limit (weak background absorption), but it is, in general, a rather poor approximation for larger $s$ because of the comparatively large contribution of nonlinear terms in $Ns$ \cite{AK15}.
To study symmetric nearest-neighbor  random walks  (the classic P\'olya walk),  we proceed from the well-known result
\beq
\label{arbdgenfunc}
P_d(z)= \frac{d}{L^d}\sum _{k _{1} =0 }^{L-1}\sum _{k_{2} =0 }^{L-1}\cdots \sum _{k _{d-1} =0 }^{L-1}\sum _{k _{d} =0}^{L-1 }
\frac{1}{d-z\,\sum_{i=1}^d C_i},
\eeq
where
\beq
C_i\equiv\cos\frac{2\pi k_i}{L}, \quad i=1,\ldots, d.
\eeq
The first term of the sum (with all the $k_i$'s equal to $0$) is the singular part, whereas the remaining contribution corresponds to the nonsingular part $Q_d(z)$.

In $d=1$ ($N=L$) one has
\beq
\label{Q1z}
Q_1(z)=\frac{1}{\sqrt{1-z^2 } } \frac{1+\left[X(z)\right]^N }{1-\left[X(z)\right]^N }-\frac{1}{N(1-z)},
\eeq
with $X(z)\equiv z^{-1} (1-\sqrt{1-z^2})$.
From here, exact results for  $\left\langle n\right\rangle $ for arbitrary values of $s$ and $N$ can be obtained \cite{AK15}:
\beq
\left\langle n\right\rangle =\frac{Ns+\sqrt{2s-s^{2} } \left(1-\frac{2}{1+\left[X(1-s)\right]^{N} } \right)}{s^{2} (N-1)} .
\eeq
In the limit of weak background absorption, the series expansion \eqref{expQs} yields
\beq
\label{wld1}
\frac{\langle n \rangle}{N} =\frac{N+1}{6} -\frac{(N-2)(N+1)\left(N+2\right)}{30} s+\mathcal{O}(s^{2}).
\eeq

For a $d=2$ dimensional walk ($N=L^2$), results for arbitrary values of $s$ were obtained for square planar lattices in the limit of small $N$ \cite{AK15}. For asymptotically-large $N$,  results for square planar lattices in the small $s$ limit have also been derived.  From Eq.\ \eqref{expQs}, one finds to leading order \cite{AK15}
\beq
\label{wld2}
\frac{\langle n \rangle}{N} =\pi ^{-1} \log N-(\pi ^{-2} \log ^{2} N+a_1' )Ns+\mathcal{O}(s^{2}),
\eeq
with $a_1'=0.061871145451\cdots$. In Sec.\ \ref{sec2}, we illustrated the Markov method for computing the mean walklength, and  the absorption probability at the deep trap for a specific example, namely the $5\times5$ periodic square planar lattice.  In the framework of the generating function formalism, the corresponding calculation proceeds as follows.  Setting $d=2$ and $N=L^2=25$ in Eq.\ \eqref{arbdgenfunc}, one has
\beq
P_2(z)= \frac{2}{25} \sum _{k_{1} =0}^{4} \sum _{k_{2} =0}^{4}\frac{1}{2-z(\cos \frac{2\pi k_{1}}{5}+\cos \frac{2\pi k_{2}}{5})}.
\eeq
Simplification of the cosines and subsequent use of Eq.\ \eqref{3.5} yields
\beq
\label{3.15}
Q_2(z)=\frac{8}{25} \left(\frac{{4}}{{ 4 + z}}+\frac{{ 16 - 6 z}}{{ 16 -12z+z}^{2} }
+\frac{4+z}{4+2z-z^{2} } \right).
\eeq
Inserting Eq.\ \eqref{3.15}  into Eq.\ \eqref{QNwl}, we recover the expression for the walklength given in Sec.\ \ref{sec2}, viz.\ Eq.\ \eqref{5b5wl}.

\section{Main results}
\label{sec4}
We now apply the methods presented in Secs.\ \ref{sec2} and \ref{sec3} to deal with the case of simple cubic (sc) lattices. The corresponding is obtained by setting $d=3$ in Eq.\ \eqref{arbdgenfunc}, i.e.,
\bal
\label{P3basic}
P_3(z)=&\frac{3}{L^3}
\sum_{k_1=0}^{L-1}\sum_{k_2=0}^{L-1}
\sum_{k_3=0}^{L-1}\frac{1}{3-{z}\left(C_1+C_2+C_3\right)}.
\eal
In \ref{app3} tabulated results for the overall mean walklength are displayed for the first few cubic lattices ($L=2,\ldots,5$). These results can be obtained either by taking Eq.\ \eqref{P3basic} as a starting point or by invoking the Markovian theory described in Sec.\ \ref{sec2}.

As one can see from the expressions in \ref{app2}, the walklength is given by ratios of polynomials of increasing degree, but it is not obvious how the resulting expressions depend on the lattice size $N$ or, equivalently, on the linear size $L$.
Were this the case, one could obtain analytic expressions similar to the formulae \eqref{wld1} and \eqref{wld2},  valid in the small $s$ limit   (or consider other limits).

The above considerations lead to the following overview of  methods for studying general problems in the theory of random walks. While the Markovian theory is very flexible and allows one to easily deal with situations such as multiple interacting walkers or imperfect lattices, the generating function approach described above is better suited for one of our main purposes, i.e., assessing the role of finite size effects in the case of a one-walker problem on a perfect periodic lattice. To this end, it will be necessary to quantify the effect of finite size corrections by computing the corresponding coefficients as accurately as possible.

The first step  consists in investigating the large $N$ behavior of the generating function $P_3(z)$ in the limit $z\to 1^-$,  which we shall develop in the following section.

\subsection{Asymptotic behavior of the lattice Green function}
\label{sec4A}

Equation \eqref{QNwl} relates the walklength $\langle n \rangle$ of a mortal walker to the nonsingular part $Q_d(z)$ of the lattice Green function. Hence, to infer the behavior of the walklength, it is of primary importance to obtain the asymptotic behavior of this quantity.

As shown by Watson \cite{W39}, the value of the nonsingular part $Q_3(z)$ in the limit when $L\to \infty$ and $z\to 1^-$ is
\bal
b_0\equiv \lim_{L\to\infty} Q_3(1)=\frac{6}{\pi^2}(\kappa+1)
K^2(\kappa),
\eal
where $\kappa\equiv [(2-\sqrt{3})(\sqrt{3}-\sqrt{2})]^2$ and
\beq
\label{elliptic}
K(k)=\int_0^1 \frac{dt}{\sqrt{(1-t^2)(1-k t^2)}}, \quad |k|<1,
\eeq
is the complete elliptic integral of the first kind. This result can be further simplified by expressing it in terms of gamma functions \cite[p.\ 614]{H95}:
\bal
b_0 &=\frac{\sqrt{6}}{32\pi^3}\Gamma\left(\frac{1}{24}\right)\Gamma\left(\frac{5}{24}\right)
\Gamma\left(\frac{7}{24}\right)\Gamma\left(\frac{11}{24}\right) \nn
&=1.5163860\cdots .
\eal
In contrast, the coefficients associated with finite size corrections to this value are,  to the best of our knowledge, unknown. Formally,  one has the following representation (see \ref{app4})
\beq
\label{Q31body}
Q_3(1)=b_0+\frac{b_1}{L}+\frac{b_3}{L^3}+\mathcal{O}(L^{-4})
\eeq
with the values $b_1=-1.354709757(1)$ and $b_3=0.2574(1)$, where the digit enclosed by parentheses is a conservative estimate of the error bar in the last digit. While these values are empirical and very accurate, it is possible to obtain a very good analytic approximation for $b_1$, with only a small contribution (less than $0.26$\%)  needing to be evaluated empirically. To the best of our knowledge,
values of the coefficients $b_1$ and $b_3$ have not been reported  in the literature.
Knowledge of these coefficients is essential to obtain accurate values for many space exploration properties of mortal and immortal P\'olya walkers on finite lattices, e.g., the mean number of distinct sites visited after a given number of steps \cite{W94}.

Even though the behavior of $Q_3(1)$ given by Eq.\ \eqref{Q31body} turns out to be sufficient to develop acceptable approximations for the walklength [cf.\ Eq.\ \eqref{av0}], the series expansion \eqref{expQs} valid for the small $s$ limit requires additional knowledge of higher order derivatives. Here, we shall restrict ourselves to study the behavior of the first-order derivative $Q_3'(1)$.
To proceed,  first we observe that there are nested relations between generating functions corresponding to different dimensionalities.  From Eq.\ \eqref{arbdgenfunc}, we obtain\footnote{Notice that the relation between generating functions corresponding to different dimensionalities via Eqs.\ \eqref{P2body} and \eqref{P3body} provides an \emph{a posteriori} justification for our non-standard notation $P_d(z)$, which makes explicit reference to the Euclidean dimension $d$. To our knowledge, such a relation does not seem to have been noticed, or to the very least emphasized, in the literature so far.}
\beq
\label{P2body}
P_2(z)=\frac{2}{L}\sum_{k_1=0}^{L-1}\frac{P_1\left(\frac{z}{2-zC_1}\right)}{2-zC_1},
\eeq
\bal
\label{P3body}
P_3(z)
=\frac{3}{(3-z)L}P_2\left(\frac{2z}{3-z}\right)+\frac{3}{L^2}\sum_{k_1=0}^{L-1}\sum_{k_2=1}^{L-1}
\frac{P_1\left(\frac{z}{3-z(C_1+C_2)}\right)}{3-z(C_1+C_2)}.
\eal

From the nested relations \eqref{P2body} and \eqref{P3body} we find
\beq
\label{Q2}
Q_2(z)=\frac{2}{(2-z)L}Q_1\left(\frac{z}{2-z}\right)+\frac{2}{L}\sum_{k_1=1}^{L-1}\frac{P_1\left(\frac{z}{2-zC_1}\right)}{2-zC_1},
\eeq
\bal
\label{Q3}
Q_3(z)
=\frac{3}{(3-z)L}Q_2\left(\frac{2z}{3-z}\right)+\frac{3}{L^2}\sum_{k_1=0}^{L-1}\sum_{k_2=1}^{L-1}\frac{P_1\left(\frac{z}{3-z(C_1+C_2)}\right)}{3-z(C_1+C_2)}.
\eal
On the other hand, the exact expression \eqref{Q1z} implies
\beq
\label{Q1}
Q_1(1)=\frac{L}{6}-\frac{L^{-1}}{6},
\eeq
\beq
Q_1'(1)=\frac{L^3}{180}-\frac{L}{9}+\frac{19L^{-1}}{180}.
\eeq
It is also known that \cite{HK82}
\beq
Q_2(1)=\frac{2}{\pi}\ln L+a_2+a_3 L^{-2}+a_4 L^{-4}+\cdots,
\eeq
\beq
Q_2'(1)=a_1' L^2-\frac{1}{\pi}\ln L+a_3'+a_4' L^{-2}+a_5' L^{-4}+\cdots,
\eeq
with
\begin{subequations}
\beq
a_2=0.195062532\cdots,\quad a_3=-0.116964779\cdots,
\quad a_4=0.484065704\cdots,
\eeq
\beq
a_1'=0.061871145451\cdots,\quad a_3'=-0.1347623119\cdots,
\eeq
\beq
a_4'=0.2005850758\cdots,\quad a_5'=0.4283683639\cdots.
\eeq
\end{subequations}

The goal now is to find (empirically) the asymptotic behavior of $Q_3'(1)$.
{A low-degree polynomial fit of the exact evaluation of $Q_3'(1)$}
up to $L=500$
yields
\begin{equation}
\label{Q3prime1}
Q_3'(1)=c_0 L+c_1+\mathcal{O}(L^{-1}),
\end{equation}
{where}
\begin{equation}
{c_0=0.381871(1),\quad c_1=-1.0785(1).}
\end{equation}
The fit also gives $c_2=0.7954(1)$ for the coefficient of $L^{-1}$ but, for the sake of caution, it will not be used in Eq.\ \eqref{Q3prime1}.

Equation  \eqref{Q31body}, taken in conjunction with Eq.\ \eqref{Q3prime1}, allows us to investigate the lattice size dependence of the walklength in the small $s$ limit for the one-walker problem [via Eq.\ \eqref{expQs}], as was already demonstrated for dimensions $d=1$ and $d=2$ in Ref.\ \cite{AK15}   [cf.\ Eqs.\ \eqref{wld1} and  \eqref{wld2}].

\subsection{One-walker problem}

\subsubsection{Exact results vs. simulation results}

The influence of competing reaction centers on the probability of reaction at a target site and on the mean walklength of the random walker before localization was studied in Ref.\ \cite{AK15}  for planar surfaces of different topologies.  See Tables I--IV in Ref.\ \cite{AK15} for a summary of analytic results and Table V for representative numerical results.

The first of the extensions considered in the present study is to expand the reaction space from $d=2$ to $d=3$.  Analytic results for the first few cubic lattices were derived using both Markov theory and generating functions, and the results presented in \ref{app2}.

\begin{table}
\caption{\label{tab1}Role of the probability $s$ in influencing reaction efficiency in the one-walker problem. The reaction space is a $d=2$, $L\times L$ square-planar lattice subject to periodic boundary conditions. The tabulated quantity is $\langle n\rangle/N$; ``Exact''  denotes  results obtained using (both) lattice Green's functions and  the theory of finite Markov processes (see text).}
\begin{tabular}{lllllll}
\hline
$L$& $s$& Exact&MC&$s$&Exact&MC\\
\hline
3 & 0 & 1 &1.0000  & 0.01 & 0.92498 & 0.9246 \\
5 & 0 & 1.2667 &1.2668  & 0.01 & 0.96590 & 0.9658 \\
7 & 0 & 1.4615 &1.4614  & 0.01 & 0.85115 & 0.8508 \\
9 & 0 & 1.6124 &1.6124  & 0.01 & 0.69643 & 0.6954 \\
11 & 0 & 1.7350 & 1.7350 & 0.01 & 0.55640 & 0.5564 \\
13 & 0 &1.8382  &1.8381  & 0.01 &0.44440  & 0.4442\\
15 & 0 &1.9271  &1.9268  & 0.01 &0.35836  &0.3583 \\
17 & 0 &2.0053  &2.0049  & 0.01 &0.29277  &0.2927 \\
19 & 0 &2.0750  &2.0744  & 0.01 &0.24247  &0.2425 \\
21 & 0 &2.1379  &2.1370  & 0.01 &0.20345  &0.2035 \\
\hline
\end{tabular}
\end{table}

To complement (and later extend) the analytic results in $d=2$, we have carried out MC simulations for a series of  $L\times L$ square-planar lattices subject to periodic boundary conditions. As in Ref.\ \cite{AK15}, the reaction center is static and positioned at the centrosymmetric site of the planar lattice, and $s$ denotes the probability of absorption of the diffusing reactant at other sites.

For each lattice, a varying number of statistical realizations were carried out sequentially to ensure that the amplitude of the $95\%$ confidence intervals for $\langle n \rangle$ was $<0.01$. Because of the reduced dispersion in the length of the random walk trajectories for larger $s$ values, the number of realizations needed to ensure the prescribed accuracy was smaller than for larger $s$ (typical values were between $3.6 \times 10^8$ and $1.5 \times 10^9$).

Displayed in Table \ref{tab1} is a comparison with results calculated using the analytic expressions derived in Ref.\ \cite{AK15} and MC simulations for two values of $s$, namely $s=0$ (no background trapping) and $s=0.01$.  The agreement between the analytic and ``experimental" results is excellent, as the relative differences are in all cases $< 10^{-3}$, and typically of the order of $10^{-4}$ or smaller.

\begin{table*}
\caption{\label{tab2}Role of the probability $s$ in influencing reaction efficiency in the one-walker problem. The reaction space is a $d=3$, $L\times L\times L$ cubic lattice subject to periodic boundary conditions. The tabulated quantity in the columns labeled ``Exact'' and ``MC'' is $\langle n\rangle/N$, where ``Exact'' denotes  results obtained using (both) lattice Green's functions and  the theory of finite Markov processes (see text). The last three columns give the percentage error of the approximations \eqref{av0}, \eqref{av1}, and \eqref{avPade}.}
\begin{tabular}{lllllll}
\hline
$s$&$L$& Exact&MC&\% Eq.\ \eqref{av0}&\% Eq.\ \eqref{av1}&\% Eq.\ \eqref{avPade}\\
\hline
0&3 & 1.1282  & 1.1282 &1.111&1.111&1.111 \\
&5 &  1.2587 & 1.2586 & 0.086&0.086 &0.086\\
&7 &  1.3276  & 1.3273 &0.013 &0.013 &0.013\\
&9 &  1.3681 & 1.3681  &3.3$\times 10^{-3}$&3.3$\times 10^{-3}$ &3.3$\times 10^{-3}$\\
&11 &  1.3945 & 1.3945  &1.1$\times 10^{-3}$&1.1$\times 10^{-3}$ &1.1$\times 10^{-3}$\\
&13 &  1.4129  &1.4129 & 4.8$\times 10^{-4}$&4.8$\times 10^{-4}$ &4.8$\times 10^{-4}$\\
&15 & 1.4266 & 1.4266 &2.3$\times 10^{-4}$ &2.3$\times 10^{-4}$ &2.3$\times 10^{-4}$\\
\hline
0.005&3 & 0.98241 &0.98239  &0.820&0.848&0.848\\
&5 & 0.70539&0.70535 &0.174&0.014&0.012\\
&7  & 0.40519 &0.40518  &0.190&4.8$\times 10^{-3}$&8.4$\times 10^{-3}$\\
&9 & 0.22845 &0.22845 &0.145&6.4$\times 10^{-4}$&3.2$\times 10^{-3}$\\
&11 & 0.13559  &0.13559 &0.107&3.5$\times 10^{-3}$&2.1$\times 10^{-4}$\\
&13 &   0.08549 & 0.08549 &0.080&4.8$\times 10^{-3}$&1.3$\times 10^{-4}$\\
&15 &0.05688 &0.05688  &0.061&5.3$\times 10^{-3}$&2.1$\times 10^{-3}$\\
\hline
0.01&3 & 0.87000 & 0.8700&0.596&0.645&0.645\\
&5 & 0.49002 & 0.4903 &0.273&0.011&0.017\\
&7  & 0.23909 & 0.2390 &0.223&2.8$\times 10^{-3}$&0.011\\
&9 & 0.12464 & 0.1246 &0.154&6.1$\times 10^{-3}$&2.2$\times 10^{-3}$\\
&11 & 0.07126  & 0.07129&0.108&9.3$\times 10^{-3}$&1.6$\times 10^{-3}$\\
&13 &   0.04408 & 0.04411 &0.078&0.010&3.3$\times 10^{-3}$\\
&15 &0.02902  &0.02902  &0.058&0.011&4.1$\times 10^{-3}$\\
\hline
%0.05&3 & 0.45431 &  &0.219&0.091&0.088\\
%&5 & 0.14238 & &0.406&0.014&0.055\\
%&7  & 0.05588 &  &0.232&0.038&8.3$\times 10^{-3}$\\
%&9 & 0.02689 & &0.139&0.047&4.3$\times 10^{-3}$\\
%&11 & 0.01486  & &0.089&0.046&8.9$\times 10^{-3}$\\
%&13 &   0.009044 &  &0.060&0.042&8.9$\times 10^{-3}$\\
%&15 &0.005901 &  &0.042&0.039&8.7$\times 10^{-3}$\\
%\hline
0.1&3 & 0.28452 & 0.28450 &0.524&0.362&0.356\\
&5 & 0.07548 &0.07553 &0.402&0.028&0.056\\
&7  & 0.02855 &0.02855  &0.210&0.085&5.4$\times 10^{-3}$\\
&9 & 0.01358 &0.01358 &0.119&0.088&5.9$\times 10^{-3}$\\
&11 & 0.007473  &0.007478 &0.073&0.081&8.4$\times 10^{-3}$\\
&13 &   0.004537 &0.004538  &0.048&0.074&8.4$\times 10^{-3}$\\
&15 &0.002957 &0.002958  &0.033&0.067&7.6$\times 10^{-3}$\\
\hline
\end{tabular}
\end{table*}

For the same set of conditions described in the preceding paragraph, MC simulations were carried out for a $d=3$ dimensional reaction space, here a $L\times L\times L$ cubic lattice subject to periodic boundary conditions. As is evident from the data in Table \ref{tab2}, agreement between results obtained from the generating function approach, the Markov approach, and MC simulations is again excellent.

\subsubsection{Small $s$ approximation}

To complement the above results, we now particularize the expansion \eqref{expQs} to the $d=3$ case. Using Eqs. \eqref{Q31body} and \eqref{Q3prime1}, we find
\beq
\label{asexpnoverN}
\frac{\langle n\rangle}{N}=b_0+\frac{b_1}{N^{1/3}}+\frac{b_0+b_3}{N}+\cdots
- \left[b_0^2-\frac{2b_0 b_1}{N^{1/3}}+\frac{b_1^2+c_0}{N^{2/3}}+\frac{b_0(b_0+2b_3)+c_1}{N}
+\cdots\right]N s +\cdots.
\eeq

\begin{figure}
\includegraphics[width=0.75\columnwidth]{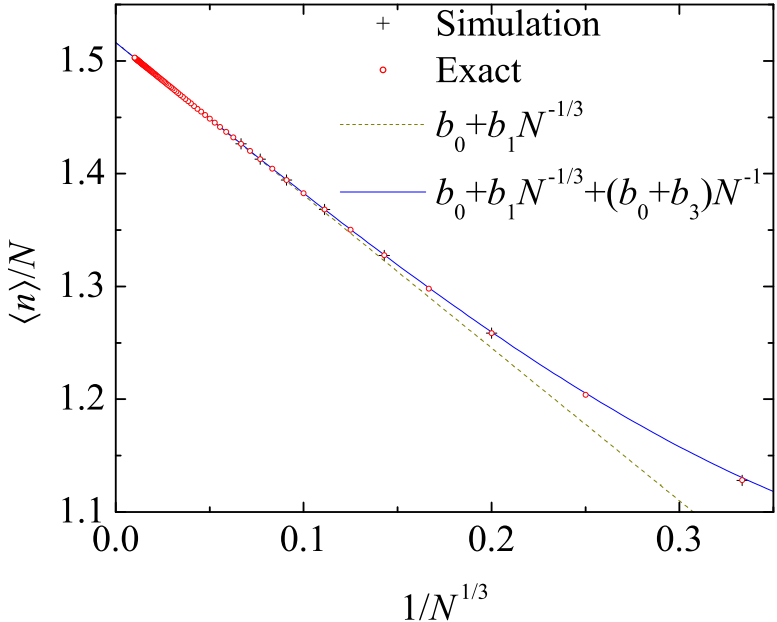}
\caption{\label{noverN3D} Comparison of exact results for $\langle n \rangle /N$ ($d=3$, $s=0$) with  approximations of decreasing order in $N$ and results from numerical simulations.}
\end{figure}

\begin{figure}
\includegraphics[width=0.75\columnwidth]{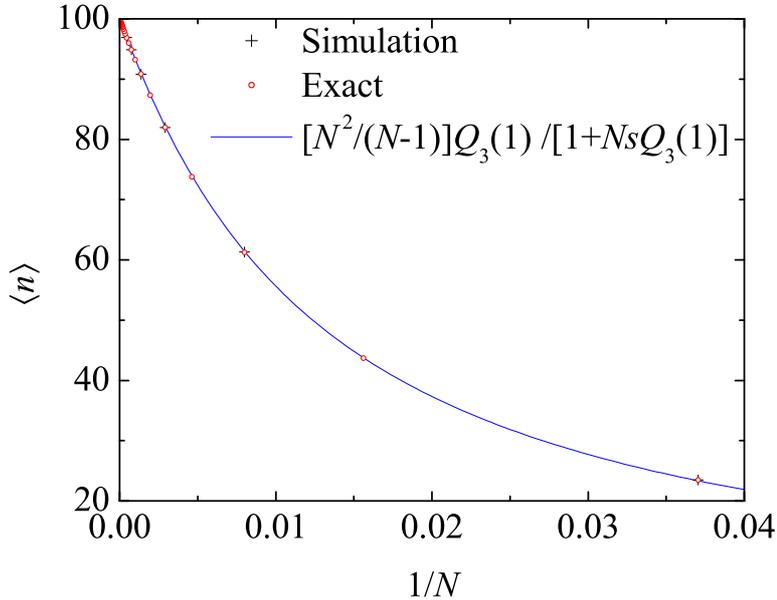}
\caption{\label{n3D} Comparison of exact results for $\langle n \rangle $ ($d=3$, $s=0.01$) with  the approximation \eqref{av0} and results from numerical simulations.}
\end{figure}

For $s=0$, one then has
\beq
\label{4.16}
\frac{\langle n\rangle}{N}=b_0+\frac{b_1}{N^{1/3}}+\frac{b_0+b_3}{N}+\cdots.
\eeq
As one can see in Fig.\ \ref{noverN3D}, approximations based on the above expansion match exact results very well.

Finally, we note that for $s=0$ the result \eqref{asexpnoverN} implies
\beq
\langle n\rangle \sim b_0 N+b_1 N^{2/3}+\mathcal{O}(1),
\eeq
which, at the level of the subdominant term, is at odds with the result originally given by Montroll, namely \cite{W69}
\beq
\langle n\rangle \sim b_0 N+\mathcal{O}(N^{1/2}).
\eeq
This discrepancy presumably arises because, in his derivation, Montroll assumed that $Q_3(z) \sim \text{const}+{\mathcal O}(\sqrt{1-z})$ in the vicinity of $z=1$. However, this is only the case for an \textit{infinite} lattice. For a finite lattice, one has $Q_3(z) \sim \text{const}+{\mathcal O}(1-z)$. In other words, $Q_3'(1)$ is finite at finite $L$, although it diverges as $Q_3'(1)\sim L$ in the limit $L\to\infty$.

\subsubsection{Alternative approximations}

An alternative approximation which works surprisingly well amounts to replacing $Q_3(1-s)$ by $Q_3(1)$ in the exact formula \eqref{QNwl} for $d=3$. One then has
\beq
\label{av0}
\frac{\langle n \rangle}{N}\approx\frac{N}{N-1}\frac{Q_3(1)}{1+Ns Q_3(1)}.
\eeq
Replacing $Q_3(1)$ with successive approximations based on the retention of  an increasingly large number of terms from the expansion \eqref{Q31body} in the above equation is expected to provide a very good approximation for sufficiently large $N$. In our calculations, we have retained all the terms on the right-hand side of Eq.\ \eqref{Q31body}, except the $\mathcal{O}(L^{-4})$ term. Figure  \ref{n3D} and Table \ref{tab2} show that  Eq.\ \eqref{av0} does extremely well, even for not so large lattices. The good performance of Eq.\ \eqref{av0} is not restricted to small $s$. In fact, we have checked that the maximum error of Eq.\ \eqref{av0} in the whole range $0\leq s\leq 1$ is about $1.1\%$, $0.41\%$, $0.24\%$, $0.15\%$, $0.11\%$, $0.08\%$,  and $0.06\%$ for $L=3$, $5$, $7$, $9$, $11$, $13$, and $15$, respectively. This shows that the  dependence of $\langle n \rangle$ on $s$ is dominated by the term $Ns$ in the denominator of Eq.\ \eqref{QNwl} rather than by the $s$-dependence of $Q_3(s)$.

For  values of $N$ even larger than those considered in Table \ref{tab2}, Eq.\ \eqref{av0} yields the expected behavior $\langle n \rangle \sim s^{-1}$. This asymptotic $s^{-1}$ law is actually universal in the sense that it does not depend on dimensionality, but rather on the fact that the probability $p$ to die at the deep trap becomes vanishingly small for sufficiently large $N$ (correspondingly, the probability to die at a any other site is almost one). A simple probabilistic argument then yields the above result \cite{WK81,WK82}.

On the other hand, it is possible to refine the approximation given by Eq.\ \eqref{av0} by taking as a starting point the Taylor-expansion of $Q_d(1-s)$:
\beq
\label{series}
Q_d(1-s)=Q_d(1)+\sum_{n=1}^\infty \frac{(-1)^n}{n!}Q_d^{(n)}(1) s^n.
\eeq
Now we particularize to $d=3$. In that case, we already know that $Q_3(1)\sim L^0$ and $Q_3'(1)\sim L$. It can also be numerically checked that $Q_3''(1)\sim L^3$. Let us \textit{conjecture} that
\beq
Q_3^{(n)}(1)\sim L^{2n-1},\quad n\geq 1.
\eeq
This conjecture is supported by the following heuristic argument. From Eqs.\ \eqref{Q2} and \eqref{Q3} one can see that one of the main contributions to $Q_3^{(n)}(1)$ comes from $Q_1^{(n)}(1)/L^2$ and, according to Eq.\ \eqref{Q1z}, $Q_1^{(n)}(1)\sim L^{2n+1}$. Under this premise, Eq.\ \eqref{series} yields
\beq
\label{series3D}
Q_3(1-s)=q_0+L^{-1}\sum_{n=1}^\infty \frac{(-1)^n q_n}{n!} (sL^2)^n,
\eeq
where
\begin{subequations}
\beq
q_0\equiv Q_3(1)\sim L^0,
\eeq
\beq
q_n\equiv \frac{Q_3^{(n)}(1)}{L^{2n-1}}\sim L^0,\quad n\geq 1.
\eeq
\end{subequations}
Then, the first few terms in the expansion in powers of $s$ of $(N-1)\langle n\rangle/N^2$ are
\bal
\frac{N-1}{N^2}\langle n \rangle=&q_0-\left(q_0^2+\frac{q_1}{L^2}\right)Ns+
\left(q_0^3+\frac{2q_0 q_1}{L^2}+\frac{q_2}{2L^3}\right)(Ns)^2-
\left(q_0^4+\frac{3q_0^2 q_1}{L^2}+\frac{q_0 q_2}{L^3}\right.\nn
&\left.+\frac{6q_1^2+q_3}{6L^4}\right)(Ns)^3+\cdots.
\eal
We observe that, at any order in $s$, the dominant and subdominant terms depend on $q_0$ and $q_1$ only. Therefore, in the limit of large $L$, we can neglect $q_n$ for $n\geq 2$ and make the approximation
\bal
\label{4.25}
\frac{N-1}{N^2}\langle n\rangle\approx &q_0\left[1-q_0 Ns+ (q_0Ns)^2 -(q_0Ns)^3+\cdots \right]-\frac{q_1Ns}{L^2}\left[1-2q_0 Ns\right.\nn
&\left.+3(q_0Ns)^2-4(q_0Ns)^3+\cdots\right]\nn
=&\frac{q_0}{1+q_0 Ns}-\frac{q_1 Ns}{L^2}\frac{1}{(1+q_0 Ns)^2}\nn
\approx&\frac{q_0-q_1 N s/L^2}{1+Ns(q_0-q_1 N s/L^2)}.
\eal
In summary, if $N\gg 1$, we can expect that
\beq
\label{av1}
\frac{\langle n \rangle}{N}\approx\frac{N}{N-1}\frac{Q_3(1)-Q_3'(1)s}{1+Ns [Q_3(1)-Q_3'(1)s]},
\eeq
with $Q_3(1)$ and $Q_3'(1)$  given, respectively, by the approximations \eqref{Q31body} and \eqref{Q3prime1}, is a reliable approximation for any $s$.
If (at finite $Ns$) one additionally neglects $q_1 N s/L^2$ versus $q_0$ in Eq.\ \eqref{4.25}, Eq.\ \eqref{av0} is recovered.
Table \ref{tab2} shows that, in general, Eq.\ \eqref{av1} indeed represents an improvement over the already very accurate approximation \eqref{av0} for $s\leq 0.1$. Nevertheless, it can be checked that, somewhat paradoxically, Eq.\ \eqref{av0} is more accurate than Eq.\ \eqref{av1} for $s>0.414$, $s>0.205$, $s>0.130$, $s>0.092$, $s>0.069$, and $s>0.054$ if $L=5$, $7$, $9$, $11$, $13$, and $15$, respectively. Essentially, the reason is that the linear approximation $Q_3(1-s)\approx Q_3(1)-Q_3'(1)s$ does not perform well when being extrapolated to the region $s\sim Q_3(1)/Q_3'(1)$ (or, equivalently, to the region $s\sim L^{-1}$).

Finally, we note that in two dimensions the approximation corresponding to Eq.\ \eqref{av0}  turns out to work very well for $s=0$ thanks to the detailed knowledge of the behavior of $Q_2(1)$ in terms of the coefficients $\{a_i\}$.  When $s>0$, the approximation remains good, but performs worse than in the $d=3$ case. A refined approximation based on the $d=2$ analog of Eq.\ \eqref{av1} yields only a very modest improvement.

A limitation of both Eqs.\ \eqref{av0} and \eqref{av1} is their inconsistency with the boundary condition $\langle n\rangle=1$ at $s=1$, which must hold regardless of the value of $N$. Such a boundary condition is equivalent to the relation $Q_d(0)=1-N^{-1}$. Obviously, neither $Q_3(1-s)\approx Q_3(1)$ nor $Q_3(1-s)\approx Q_3(1)-Q_3'(1)s$ satisfies this requirement. This shortcoming prompts one to consider the introduction of a two-point Pad\'e approximant for $Q_3(1-s)$ that satisfies the boundary condition $Q_3(0)=1-N^{-1}$ while simultaneously reproducing $Q_3(1)$ and $Q_3'(1)$ . Such an approximant reads
\beq
\label{Pade}
Q_3^\pad(1-s)=\frac{Q_3(1)(1-s)+(1-N^{-1})B s}{1-s+B s},
\eeq
where
\beq
B\equiv \frac{Q_3'(1)}{Q_3(1)-1+N^{-1}}.
\eeq
The associated approximation for $\langle n\rangle$ is then
\beq
\label{avPade}
\frac{\langle n \rangle}{N}\approx\frac{N}{N-1}\frac{Q_3^\pad(1-s)}{1+Ns Q_3^\pad(1-s)}.
\eeq
Table \ref{tab2} shows that Eq.\ \eqref{avPade} performs excellently well, generally outperforming both Eqs.\ \eqref{av0} and \eqref{av1}, especially for large $N$ and sufficiently large $s$. Given that Eq.\ \eqref{avPade} does not require more coefficients than those already present in Eq.\ \eqref{av1}, it is our recommended approximation.
{In fact, we have checked that Eq.\ \eqref{Pade} is more accurate than higher-order Pad\'e approximants that can be constructed by using the exact property $Q_d'(0)=-N^{-1}$.}

{A test of the approximations \eqref{Pade}--\eqref{avPade} for $d=1$ and $d=2$ shows a very good performance as well, although inferior to the three-dimensional case. For instance, if $s=0.1$ and $L=15$, the error in the analogues of Eq.\ \eqref{avPade} is $0.04\%$ and $0.11\%$ for $d=1$ and $d=2$, respectively.}

\subsection{Two-walker problem}

The above results are for a \textit{single} diffusing reactant and a static co-reactant (reaction center), with the remaining sites of the reaction space either passive ($s=0$) or activated ($s\neq 0$). We now suppose that the co-reactant is no longer immobile, implying that both reactant and co-reactant can be viewed as two identical walkers performing synchronous displacements between nearest-neighbor sites at each time step.  Here, attempts to exchange the position of both walkers become possible, and one must specify what happens in this case. To account for excluded volume effects, we shall assume (as in previous references \cite{KNN00,NKN01}) that both walkers are instantaneously reset to their previous positions. The only case in this genre for which exact numerical results are available corresponds to $d=2$ and no background trapping ($s=0$) (see Refs.\  \cite{KNN00,NKN01}).

\begin{table}
\caption{\label{tab3}Role of the probability $s$ in influencing reaction efficiency in the two-walker problem. The reaction space is a $d = 2$, $L\times L$ square-planar lattice subject to periodic boundary conditions.  The tabulated quantity is $\langle n \rangle/N$. For $s=0.01$, ``MC I (II)'' denotes MC simulation results for process I (II). ``Markov" denotes values obtained from Markov theory.}
\begin{tabular}{lllllll}
\hline
$L$&$s$&Markov&MC&$s$&MC I&MC  II\\
\hline
3 & 0 & 0.8889 & 0.8891 & 0.01 & 0.7805 & 0.8925 \\
5 & 0 & 1.0429 & 1.0436 & 0.01 & 0.6943 & 1.0173 \\
7 & 0 & 1.1561 & 1.1559 & 0.01 & 0.5463 & 1.0004 \\
9 & 0 & 1.2435 & 1.2435 & 0.01 & 0.4145 & 0.8887 \\
11 & 0 & 1.3129 & 1.3128 & 0.01 & 0.3150 & 0.7503 \\
\hline
\end{tabular}
\end{table}

\begin{table*}
\caption{\label{tab4}Role of the probability $s$ in influencing reaction efficiency in the two-walker problem. The reaction space is a $d = 3$, $L\times L\times L$ cubic lattice subject to periodic boundary conditions. The tabulated quantity is $\langle n \rangle/N$.``MC I (II)'' denotes MC simulation results for process I (II).}
\begin{tabular}{llllllllllll}
\hline
$L$&$s$&MC&$s$&MC I&MC II&$s$&MC I&MC II&$s$&MC I&MC II\\
\hline
3 & 0 & 0.9826 & 0.005 &0.7830&0.9832&0.01 & 0.6513 & 0.9453&0.1 &0.1678&0.3925\\
5 & 0 & 1.0693 & 0.005 &0.4599&0.8613&0.01 & 0.2935 & 0.6597&0.1 &0.04077&0.1096\\
7 & 0 & 1.1126 & 0.005 &0.2318&0.5519&0.01 & 0.1297 & 0.3494&0.1 &0.01517&0.04185\\
9 & 0 & 1.1379 & 0.005 &0.1228&0.3261&0.01 & 0.06497 & 0.1879&0.1 &0.00718&0.01997\\
11 & 0 & 1.1544 &0.005 &0.07072&0.1977& 0.01 & 0.03652 & 0.1089&0.1 &0.00394&0.01010\\
\hline
\end{tabular}
\end{table*}

We present in Tables \ref{tab3} and \ref{tab4} a MC study for the case of two co-reactants diffusing, without and with background trapping.  The prescribed simulation accuracy was the same as in the one-walker case (amplitude of the $95\%$ confidence intervals $<0.01$). Our study is restricted to odd lattices, since on an even lattice the two walkers never meet for certain initial conditions, implying that the joint walk never terminates in the $s=0$ case.

In the two-walker case, application of the generating function approach is complicated by the fact that in the comoving frame defined by one of the walkers the other walker performs a random walk with varying step length. In addition, this random walk becomes inhomogeneous as a result of the walkers' hard core interaction when attempting to exchange positions. However, as noted in our earlier remarks, the theory of finite Markov processes can still be mobilized to obtain analytic results. In this case, the transient states of the corresponding Markov chain are not specified by the position of the walker with respect to an immobile site (as in the one-walker problem), but rather by the relative position of one walker with respect to the other.

The analytic calculations given in Refs.\ \cite{KNN00,NKN01} for the square planar lattice and $s=0$ were performed according to the above procedure. Comparison between the numerically exact results obtained therein and MC results (respectively labeled ``Markov" and ``MC" in Table \ref{tab3}) is again excellent (the relative difference is of the order of $10^{-3}$ or less). These results are important, since they provide a level of confidence in simulations, which can then be used reliably to extend the results to more general situations.

As already mentioned, to our knowledge there are no analytic results available for the two-walker problem in $d=2$ reaction spaces with $s>0$, nor any results in $d=3$  for any $s$. Simulation results for special cases in this parameter regime are shown in Tables  \ref{tab3} and \ref{tab4}. Two types of situations are considered: process I, in which the joint walk is instantaneously terminated if the two walkers meet at the same lattice site or if \textit{either} of the walkers dies at any site prior to encounter; and process II, in which the walk is terminated if the two walkers meet at the same lattice site or if both of them are trapped at any site prior to encounter.

\section{Discussion}
\label{sec5}

As reviewed in Ref.\ \cite{AK15}, a myriad of physical problems provide experimental realizations of the consequences of background trapping on the efficiency of reaction between a pair of randomly diffusing co-reactants localized on a $d=2$ template.  In the context of, say, heterogeneous catalysis,  the question is whether, and to what extent, degradation owing to catalyst poisoning (in our case, a decrease in the probability $s$) has the same influence when the dimensionality of the reaction space increases from $d=2$ to $d=3$.  An exact comparison is possible only for lattices in $d=2$ and $d=3$ with the same number $N$ of lattice sites.  Here, for the square-planar and simple cubic lattices used to model the reaction space, we can analyze one lattice pair for which this constraint is nearly satisfied.

For the $11\times11$ square-planar lattice, the number of lattices sites ($121$) is close to the $125$ sites defining the $5\times5\times5$ cubic lattice.  Geometrically, expanding the reaction space from $d=2$ to $d=3$ puts more background sites closer to the static target molecule; thus, for $s=0$, the mean walklength in $d=2$, $\langle n \rangle =  209.9$,  is larger than in $d=3$,  $\langle n \rangle = 157.3$. If one now switches from $s=0$ to $s=0.01$ (see Tables \ref{tab1} and \ref{tab2}), the mean walklength decreases by $68\%$ in $d=2$ and by $61\%$ in $d=3$.  The example suggests that when background sites are activated, their relative influence on the reaction efficiency is greater in $d=2$ than in $d=3$. This conjecture is confirmed by the following argument. For $d=2$ the walklength grows more rapidly with $N$ as for $d=3$ ($\sim N \log N$ vs. $\sim N$), and so the walklength becomes larger in $d=2$.  However, activating background trapping tends to negate geometry-induced differences in the reaction efficiency for sufficiently large $N$, and the walklength drops to a value close to $s^{-1}$ both in $d=2$ and in $d=3$. This means that the decrease with respect to the value of $\langle n \rangle$ for the case where background trapping is absent is more pronounced in the two-dimensional case. For $d=1$, one has $\langle n \rangle \sim N^2$ when $s=0$ \cite{W69}, implying that activation of background trapping is expected to have an even stronger impact than in $d=2, 3$.

A comparison similar to the above one can be made for the two-walker problem.  As reported in Tables \ref{tab3} and \ref{tab4} (see MC I), the mean walklength before the reactive process terminates is $\langle n \rangle = 158.9$ in $d=2$ and $\langle n \rangle = 133.7$ in $d=3$ for the $s=0$ case, a result in qualitative agreement with the behavior noted in the previous paragraph.  When background trapping is turned on ($s=0.01$), the mean walklength decreases by $76\%$ in $d=2$ and by $72.6\%$ in $d=3$. For the process II, the relative decrease in the walklength is also higher in $d=2$ (cf.\ MC II in Tables \ref{tab3} and \ref{tab4}). Thus, for both variants, I and II, of the two-walker problem, we arrive at the same conclusion as for the one-walker case.

Having analyzed the influence of a change in $s$ in different spatial dimensions, we now turn to the study of the comparative efficiency of the one-walker problem (first-order kinetics) and the two-walker problem (second-order kinetics) in different spatial dimensions, both with and without background trapping ($s=0.01$ and $s=0$, respectively). For the two-walker problem, taking again the cases of the $5\times5\times 5$ lattice and the $11\times11$ as a reference, one finds that the value of the walklength is in all cases smaller in $d=3$ than in $d=2$, i.e., the reaction is more efficient in three dimensions. While this was already known to be true for the one-walker problem, here it is also corroborated for the two-walker problem.

Finally, we note that, irrespective of the value of $s$, when the second co-reactant becomes mobile, the relative decrease in the walklength is larger in $d=2$ than in $d=3$. In lower dimensions mixing effects due to enhanced diffusion have a stronger impact on the efficiency of the underlying diffusion-controlled process, since the geometric constraints imposed by the template are more severe, and mobility is important to increase the rate of reactive collisions.

Let us now discuss possible implications of our findings for a simple model of catalyst activation.  Assume that the lattice represents a catalytic substrate with active sites subject to inhomogeneous degradation. The case $s=1$ would correspond to a perfect catalyst, whereas the case $s<1$ would imply that all the sites but one suffer partial deactivation. Local differences in the degree of activation of individual catalytic sites (selective poisoning) may arise owing to differences in affinity of  the chemisorbed poison species \cite{G11}. In the present case, if one takes a perfect catalyst as a starting point ($s=1$), as catalyst degradation progresses, one initially has universal $1/s$ behavior in $d=2$ and $d=3$, but then the trapping time increases faster in $d=2$, implying that a $d=3$ lattice is more robust against the deactivation process.

While the type of inhomogeneous catalyst degradation considered here may seem rather specific, we point out that, due to the translational invariance of the lattice, the results for the one-walker case are more general than what they might look at first sight since (a) they do not depend on the specific location of the lattice site immune to poisoning (deep trap), which may actually be chosen at random, and (b) for a lattice of a given size containing a deep trap, the result also holds for any lattices constructed by juxtaposing $n$ replicas of the original lattice (each of them containing one deep trap). Note also that, according to the scheme \eqref{sch}, if one imposes confining rather than periodic boundary conditions on an odd lattice, our results continue to hold, since any time the walker attempts to leave the lattice, it bounces back to the same site, implying that the symmetry class is conserved (as is the case for the periodic lattice).

The above results also emphasize that the simplest inhomogeneity in the catalyst deactivation process suffices to induce interesting, nonlinear behavior. This finding acquires special relevance when considering the recent, rapidly accelerating progress in the design of nanocatalysts at the level of their individual constituents \cite{FDPFB15, CDS12}. If the site degradation probability $1-s$ is assumed to be proportional to the concentration $c$ of the poisoning agent in a certain regime, the asymptotic $1/s$ behavior of the walklength (corresponding to homogeneous, non-selective poisoning) results in a linear decrease of the ``reaction rate'' $1/\langle n  \rangle$ quantifying the catalytic efficiency as a function of $c$. However, a single site immune to poisoning suffices to destroy this linear dependence, resulting in dramatic deviations for increasingly small systems (decreasing $N$). For instance, under the above assumption, one has $s=1-\mu c$, where $\mu$ is a suitably chosen proportionality constant. Inserting this ansatz into the expressions given in \ref{app2}, the nonlinear behavior in $c$ can be characterized in more detail for the $d=3$ case. Taken together with the results in Ref.\ \cite{AK15}, the general conclusion reaffirms that the observed nonlinearities are very sensitive to the geometric details of the support, and notably to dimensionality and finite size effects.

A refinement of the present model may unveil interesting features for systems where selective poisoning takes place and a nonlinear dependence of the catalytic activity on $c$ are observed (e.g., catalytic reactions facilitated by acid zeolites \cite{G11}, or the conversion of para-$H_2$ on a Pt-foil in the presence of CO poisoning \cite{VH74}).

Finally, let us turn our attention to  the problem of exciton trapping in photosynthetic units. A basic model considered by Montroll \cite{W69} is inspired by the idea that chlorophyll molecules form some kind of regular lattice in which exciton traps are embedded. Each chlorophyll molecule has the same \emph{a priori} probability to capture a photon and to convert it into an excitation which performs a lattice walk until it reaches a (static) reaction center. This trapping event then triggers production of sugar and carbohydrates. Here, two scenarios are possible.

First, all  lattice sites may be regarded as reaction centers, one of which is fully activated, with the remaining $N-1$ sites partially so. Here, turning on the trapping at background sites increases the efficiency of the photosynthetic event.  Even for small values of $s$, these imperfect traps \emph{increase} dramatically the rate $\langle n \rangle^{-1}$ of the photosynthetic process. Here, increasing $s$ becomes advantageous for photosynthesis, as is also the case for the catalyst system.

Conversely, the reaction site is assumed to be a perfect trap, whereas all other sites are considered to be identical imperfect traps with absorption probability $s$ (the case elaborated in \cite{AK15} and referenced in Sec.\ \ref{sec1}).  These traps may ``kill" the exciton prematurely, i.e., before it is trapped by the reaction center.   Even for small values of $s$, these imperfect traps reduce dramatically the probability $p = 1-s \langle n \rangle$ of the excitation arriving at the reaction center. Thus, increasing $s$ is a disadvantage for photosynthesis, as opposed to the catalyst system, where the reaction rate grows when $s$ becomes larger.

For a given value of $s$ and a fixed value of $N$, the walklength is larger in $d=2$ than in $d=3$, therefore $p$ remains larger in $d=3$ over the whole range of $s$. However, differences become much less pronounced in the limits of large $N$ and/or $s$ close to 1, since $\langle n \rangle\to 1/s$ in both cases.

\section{Conclusions and Outlook}
\label{sec6}

In this work, we have extended previous calculations for the walklength of a mortal P\'olya walker on periodic lattices in $d=1$ and $d=2$ to the case of the sc lattice $d=3$. We have also relaxed the original restriction of a static co-reactant by allowing for the possibility that both co-reactants perform (mortal) P\'olya walks. For lattices with similar values of $N$, the relative decrease in the walklength when background trapping is turned on is larger the lower the spatial dimension. Inasmuch as the analytic difficulties encountered in seeking a closed expression for the walklength in terms of the lattice size are extremely challenging, we have developed and assessed the worth of several approximations, notably one based on a two-point Pad\'e approximant which works surprisingly well. In our route to these results, we have been led to compute semiempirically the first finite size corrections to the sc lattice Green function. These results will hopefully pave the way for the study of many space exploration properties of random walks in sc lattices. A first consequence is the already noted discrepancy at the level of the first subdominant term between our result for the walklength on the sc lattice and the corresponding result reported earlier by Montroll for the $s=0$ case \cite{W69}.

A possible extension of the results for the one-walker problem to dimensions $d>3$  would in principle lead to the cumbersome task of computing generating functions involving an increasing number of nested sums. While this problem is beyond the scope of the present work, we have derived (see \ref{app5}) an alternative, integral representation of $P_d(z)$ which might prove useful to investigate the behavior of the walklength in higher dimensions, viz.,
\beq
\label{intrepP}
P_d(z)=\frac{d}{z} \int _{0}^{\infty }d\zeta\,e^{-\zeta d/z} \left(\frac{1}{L} \sum _{k=0}^{L-1}e^{\zeta \cos (2\pi k/L)}  \right)^{d} .
\eeq

A second extension of the present work would consist in computing  higher-order moments of the walklength for the one-walker problem. The moments can be expressed in terms of higher order $z$ derivatives of $P_d(z)$. For small enough values of $s$, the initial-condition-averaged variance
\beq
\frac{1}{N-1}\sum_{\vec{\ell }\ne \overrightarrow{0}}
\left(\langle n^2 \rangle_{\vec{\ell}}-\langle n \rangle_{\vec{\ell}}^2\right)
\eeq
is expected to be large (actually, of the order of $\langle n \rangle^2$ itself \cite{A05}), especially because of the contributions from sites $\vec{\ell}$ which are far away from the origin $\overrightarrow{0}$. This reflects the large trajectory-to-trajectory scattering observed in different realizations of the random walk. Hence, one anticipates that a modest increase in $s$  will decrease this variability drastically.

We also note that the generating function method can be used  to compute the mean walklength $\langle n \rangle_{\overrightarrow{0}}$ (and higher order moments) for the case where the walker is allowed to start at the static deep trap, and is only absorbed the first time it revisits the trap (provided that it has not been killed before as a result of background trapping). While a detailed study is beyond the scope of the present work,  it is straightforward to derive  the identity $\langle n \rangle_{\overrightarrow{0}}=[s P_d(1-s)]^{-1}$, from which one would proceed to determine the moments.

These are of course only some possibilities out of many space exploration properties that can be extracted from $Q_d(z)$ and the associated higher-order $z$-derivatives evaluated at $z=1-s$. Examples include the mean number of distinct sites visited after a given number of time steps (of special relevance for rate theories of diffusion controlled reactions), the average number of sites revisited at least or exactly a given number of times, the average number of returns to the origin, etc. \cite{YAL13,YAL14}.

To extend the present work, one could also consider more complicated settings involving (a) inhomogeneous background trapping, whereby each site $i$ is assigned a different trapping probability $s_i$, and (b) inhomogeneous diffusion, where the mobility of the particle/excitation at certain sites is decreased drastically as a result of large activation barriers or absorption processes followed by reemission, say. Of special interest in this context is the computation of the probability that the walk is terminated at a given site (splitting probability), as well as of the associated conditional mean walklength \cite{CBM07, BV14}.

In both situations (a) and (b), the translational invariance of the lattice is broken, and one must deal with defective lattices. When a translationally invariant lattice is perturbed by introducing a (very) small number of defects (trapping sites), generating function approaches can still be used \cite{W64,EN85}. However, with increasing number of trapping sites such approaches quickly becomes impractical. Up to some exceptions, a similar remark applies to the case of boundary conditions other than periodic, often used to describe rather common experimental situations, such as confining or open boundaries. In all of such cases, it is preferable to apply Markov theory in tandem with MC simulations.

Summarizing, we have shown that three different approaches, viz.\ the Markov method, the generating function method, and MC simulations, can be mobilized to obtain results of interest in the lattice theory of diffusion-controlled chemical reactions and/or physical processes. The results obtained using this triangulated approach mutually reinforce and complement each other. Our results suggest that the theoretical insights obtained can serve to cast light on two complex problems, i.e., diffusion on a partially poisoned catalytic substrate and photosynthetic trapping of excitations.  A more realistic application of our three-fold approach to either problem is possible, but would require a full-length study,  one in which there is a concrete interface with existing experimental data.

\section*{Acknowledgments}
E.A. and A.S. acknowledge the financial support of the Spanish Agencia Estatal de Investigación through Grant No. FIS2016-76359-P and of the Junta de Extremadura (Spain) through Grant No. GR18079, both partially financed by ``Fondo Europeo de Desarrollo Regional'' funds.

\appendix

\section{Site specific walklengths for the $5\times5$ lattice}
\label{app1}
\beq
\langle n \rangle_ 1 =8\frac{15 + 23 s - 3 s^2 - 3 s^3}{5 + 161 s + 158 s^2 - 50 s^3 -
 19 s^4 + s^5},
 \eeq
\beq
\langle n \rangle_ 2=2
\frac{75 + 94 s - 28 s^2 - 14 s^3 +  s^4}{5 + 161 s + 158 s^2 -
 50 s^3 - 19 s^4 + s^5},
 \eeq
\beq
\langle n \rangle_ 3 =16
\frac{10 + 11 s - 4 s^2 -  s^3}{5 + 161 s + 158 s^2 - 50 s^3 -
 19 s^4 + s^5},
 \eeq
\beq
\langle n \rangle_ 4=2
\frac{85 + 80 s - 30 s^2 - 8 s^3 +  s^4}{5 + 161 s + 158 s^2 -
 50 s^3 - 19 s^4 + s^5},
\eeq
\beq
\langle n \rangle_ 5=4
\frac{45 + 33 s - 9 s^2 - 5 s^3}{5 + 161 s + 158 s^2 - 50 s^3 -
 19 s^4 + s^5}.
\eeq

\section{Derivation of absorption probabilities at the deep trap}
\label{app2}
 The probability that a mortal walker starting at site $\vec{\ell } \neq \vec{0}$ hits the deep trap located at $\vec{0}$ exactly after $n$ steps is
\cite{AK15}
\beq
p_{T,n}(\vec{\ell })=(1-s)^n \FP_n(\vec{\ell }),
\eeq
where $\FP_n(\vec{\ell })$ is the probability to hit the deep trap in the absence of background trapping. The above equation follows directly from the notion of conditional probability. In terms of the generating function $\FP(\vec{\ell },z)\equiv\sum_{n=0}^\infty \FP_n(\vec{\ell }) z^n$ one has
\beq
\label{EqpTn2}
p_{T,n}(\vec{\ell})=\frac{(1-s)^n}{n!}\frac{\partial^n}{\partial z^n} \left. \FP(\vec{\ell},z) \right|_{z=0}
\eeq
On the other hand, one has the well-known relation \cite{MW65}
\beq
\label{EqFz}
\FP(\vec{\ell},z)=\frac{\PP(\vec{\ell},z)}{\PP(\vec{0},z)},
\eeq
where
\beq
\PP(\vec{\ell},z)=\sum _{n=0}^{\infty}\PP_n(\vec{\ell}) z^{n}.
\eeq
Inserting Eq.\ \eqref{EqFz} into Eq.\ \eqref{EqpTn2}, averaging the resulting equation over the $N-1$ possible locations of the starting site $\vec{\ell}$, and taking into account the probability conservation relation $\sum_{\vec{\ell}\neq \vec{0}}\PP(\vec{\ell},z)+\PP(\vec{0},z)=(1-z)^{-1}$, Eq.\ \eqref{EqpTn} for the initial-condition-averaged probability $p_{T,n}=(N-1)^{-1} \sum_{\vec{\ell}\neq \vec{0}} p_{T,n}(\vec{\ell})$ follows, where the short-hand notation $P_d(z)\equiv \PP(\vec{0},z)$ has been used.

\section{Overall walklengths for sc lattices of increasing size}
\label{app3}

\subsection{$2\times2\times2$}

\beq
\langle n \rangle=\frac{1}{7}\frac{116-52s-s^2}{2+14s-7s^2}.
\eeq

\subsection{$3\times3\times3$}

\beq
\langle n \rangle= \frac{18}{13}\frac{22+7s-3s^2}{1+30s+9s^2-4s^3}.
\eeq

\subsection{$4\times4\times4$}

\beq
\langle n \rangle= \frac{4}{63}\frac{6068+4772s-7093s^2+952s^3+484s^4-80s^5}{
5+286s+291s^2-444s^3+61s^4+30s^5-5s^6}.
\eeq

\subsection{$5\times5\times5$}

\begin{align}
\langle n \rangle= &\frac{1}{31}\left(1341250 + 5413125 s + 4313750 s^2 - 1113875 s^3 - 1444050 s^4 +
 65975 s^5 \right.\nonumber\\
 &\left.+ 109450 s^6 - 6825 s^7 - 784 s^8\right)/\left(275 + 44250 s + 174575 s^2 + 137815 s^3 - 36195 s^4\right.\nonumber\\
 &\left. - 46213 s^5 +
 2173 s^6 + 3501 s^7 - 220 s^8 - 25 s^9\right).
\end{align}

\section{Derivation of finite size corrections to $Q_3(1)$}
\label{app4}
\subsection{Formally exact expressions for $Q_3(1)$}
{}From the nested relations \eqref{Q2}, \eqref{Q3}, and from Eq.\ \eqref{Q1} one finds for $z=1$
\beq
\label{Q21}
Q_2(1)=\frac{2}{L}Q_1(1)+R_2,
\eeq
\bal
\label{Q31}
Q_3(1)
=&\frac{3}{2L}\left[Q_2\left(1\right)+R_2\right]+R_3\nn
=&\frac{3}{L}Q_2\left(1\right)-\frac{3}{L^2}Q_1(1)+R_3,
\eal
where we have introduced the quantities
\beq
\label{R2}
R_2\equiv\frac{2}{L}\sum_{k_1=1}^{L-1}\frac{P_1\left(\frac{1}{2-C_1}\right)}{2-C_1},
\eeq
\beq
\label{R3}
R_3\equiv\frac{3}{L^2}\sum_{k_1=1}^{L-1}\sum_{k_2=1}^{L-1}\frac{P_1\left(\frac{1}{3-C_1-C_2}\right)}{3-C_1-C_2}.
\eeq

Equations \eqref{3.5} and \eqref{Q1z} allow one to perform the decomposition $P_1(z)=P_{1A}(z)+P_{1B}(z)$ with
\beq
P_{1A}(z)\equiv\frac{1}{\sqrt{1-z^2}}, \quad P_{1B}(z)\equiv\frac{1}{\sqrt{1-z^2}}\frac{2\left[X(z)\right]^L}{1-\left[X(z)\right]^L}.
\eeq
Consequently, $R_2$ and $R_3$ may also be decomposed as $R_2=R_{2A}+R_{2B}$ and $R_3=R_{3A}+R_{3B}$, respectively.
In particular,
\begin{subequations}
\beq
\label{R2A}
R_{2A}=\frac{1}{L}\sum_{k_1=1}^{L-1}F_0\left(\frac{\pi k_1}{L},0\right),
\eeq
\beq
\label{R3A}
R_{3A}=\frac{3}{2L^2}\sum_{k_1=1}^{L-1}\sum_{k_2=1}^{L-1}F_0\left(\frac{\pi k_1}{L},\frac{\pi k_2}{L}\right),
\eeq
\end{subequations}
where
\beq
F_0(x_1,x_2)\equiv
\frac{1}{\sqrt{\sin^2 x_1+\sin^2x_2}} \frac{1}{\sqrt{1+\sin^2x_1+\sin^2x_2}}.
\eeq

Combining Eqs.\ \eqref{Q21} and \eqref{Q31},
\beq
\label{Q3(1)}
Q_3(1)=\frac{3}{L^2}Q_1(1)+T_{3A}+\frac{3}{L}R_{2B}+R_{3B},
\eeq
where
\bal
\label{C12}
T_{3A}\equiv& \frac{3}{L}R_{2A}+R_{3A}\nn
=&\frac{3}{2L^2}\sum_{k_1=0}^{L}\sum_{k_2=1}^{L-1}F_0\left(\frac{\pi k_1}{L},\frac{\pi k_2}{L}\right).
\eal
Now we consider the asymptotic properties for $L\gg 1$.

\subsection{Asymptotic behavior of $R_{2B}$ and $R_{3B}$}
Let us write $R_{2B}$ [cf.\ Eq.\ \eqref{R2}] as
\beq
\label{R2B}
R_{2B}=\frac{4}{L}\sum_{k_1=1}^{[L/2]}\frac{P_{1B}\left(\frac{1}{2-C_1}\right)}{2-C_1},
\eeq
where $[L/2]$ stands for the integer part of $L/2$ and we have taken into account that $\cos(2\pi-x)=\cos x$. We now follow Montroll's method \cite{W69}, according to which only the first few terms are relevant. Let us define
\beq
z_{1}\equiv \frac{1}{2-\cos\frac{2\pi k_1}{L}}.
\eeq
Then, for small $k_1/L$,
\begin{subequations}
\beq
\frac{z_{1}}{\sqrt{1-z_{1}^2}}=\frac{L}{2\pi k_1}-\frac{\pi k_1}{6L}+\cdots ,
\eeq
\beq
X(z_{1})=1-\frac{2\pi k_1}{L}+\frac{2\pi^2 k_1^2}{L^2}+\cdots,
\eeq
\beq
\left[X(z_{1})\right]^L=e^{-2k_1\pi}\left(1+\frac{2\pi^3 k_1^3}{3L^2}-\frac{\pi^5 k_1^5}{3L^4}+\cdots\right).
\eeq
\end{subequations}
Therefore,
\beq
\label{RR2B}
R_{2B}=r_{2B}^{(0)}+\frac{r_{2B}^{(2)}}{L^2}+\mathcal{O}(L^{-4}),
\eeq
where the dominant contribution in the large size limit
\beq
r_{2B}^{(0)}=\frac{4}{\pi}\sum_{k_1=1}^\infty\frac{1}{k_1}\frac{e^{-2\pi k_1}}{1-e^{-2\pi k_1}}=0.00238437\cdots
\eeq
is obtained by taking $L\to\infty$ in the upper limit of the sum in Eq.\ \eqref{R2B}. Further, one has
\bal
r_{2B}^{(2)}=&\frac{4\pi}{3}\sum_{k_1=1}^\infty{k_1}\frac{e^{-2\pi k_1}}{1-e^{-2\pi k_1}}\left(\frac{2\pi k_1}{1-e^{-2\pi k_1}}-1\right)\nn
=&0.04183562895\cdots.
\eal

Analogously, in the case of
\beq
R_{3B}=\frac{12}{L^2}\sum_{k_1=1}^{[L/2]}\sum_{k_2=1}^{[L/2]}\frac{P_{1B}\left(\frac{1}{3-C_1-C_2}\right)}{3-C_1-C_2}
\eeq
we define
\beq
z_{12}\equiv \frac{1}{3-\cos\frac{2\pi k_1}{L}-\cos\frac{2\pi k_2}{L}}.
\eeq
Then, for small $k_1/L$ and $k_2/L$,
\begin{subequations}
\beq
\frac{z_{12}}{\sqrt{1-z_{12}^2}}=\frac{L}{2\pi k}-\frac{\pi}{6L}\frac{k^4+k_1^2k_2^2}{k^3}+\cdots ,\quad k\equiv\sqrt{k_1^2+k_2^2},
\eeq
\beq
X(z_{12})=1-\frac{2\pi k}{L}+\frac{2\pi^2 k^2}{L^2}+\cdots,
\eeq
\bal
\left[X(z_{12})\right]^L=e^{-2\pi k}\left[1+\frac{2\pi^3 (k^4-k_1^2k_2^2)}{3k L^2}-\frac{\pi^5(15 k^8-16 k^4 k_1^2k_2^2-5k_1^4k_2^4)}{45k^3L^4}+\cdots\right].
\eal
\end{subequations}
Consequently,
\beq
\label{RR3B}
R_{3B}=\frac{r_{3B}^{(1)}}{L}+\frac{r_{3B}^{(3)}}{L^3}+\mathcal{O}(L^{-5}),
\eeq
with
\begin{subequations}
\beq
r_{3B}^{(1)}=\frac{12}{\pi}\sum_{k_1=1}^\infty\sum_{k_2=1}^\infty\frac{1}{k}\frac{e^{-2\pi k}}{1-e^{-2\pi k}}=0.000376447\cdots,
\eeq
\bal
r_{3B}^{(3)}=4{\pi}\sum_{k_1=1}^\infty\sum_{k_2=1}^\infty\frac{1}{k^2}\frac{e^{-2\pi k}}{1-e^{-2\pi k}}\left[2\pi\frac{k^4-k_1^2k_2^2}{1-e^{-2\pi k}}-\frac{k^4+k_1^2k_2^2}{k}\right]=0.0138005442\cdots.
\eal
\end{subequations}
\subsection{Asymptotic behavior of $T_{3A}$}
Again, we closely follow Montroll's approach \cite{W69}, originally devised for $R_{2A}$. First, we define the function
\bal
\label{C23}
F(x_1,x_2)\equiv&
F_0(x_1,x_2)-\frac{1}{\sqrt{x_1^2+x_2^2}}-\frac{1}{\sqrt{(\pi-x_1)^2+x_2^2}}\nn
&-\frac{1}{\sqrt{x_1^2+(\pi-x_2)^2}}-\frac{1}{\sqrt{(\pi-x_1)^2+(\pi-x_2)^2}}
\eal
and decompose $T_{3A}$ as follows,
\beq
\label{T3A}
T_{3A}=T_{3A,1}+T_{3A,2},
\eeq
where [cf.\ Eq.\ \eqref{C12}]
\begin{subequations}
\beq
\label{C24}
T_{3A,1}\equiv\frac{3}{2L^2}\sum_{k_1=0}^{L}\sum_{k_2=1}^{L-1}F\left(\frac{\pi k_1}{L},\frac{\pi k_2}{L}\right),
\eeq
\beq
\label{C25}
T_{3A,2}
\equiv\frac{6}{\pi L}\sum_{k_1=0}^{L}\sum_{k_2=1}^{L-1}\frac{1}{\sqrt{k_1^2+k_2^2}}.
\eeq
\end{subequations}

\subsubsection{Euler--Maclaurin formula}
According to the Euler--Maclaurin summation formula \cite{AS72},
\bal
\label{EM}
\frac{1}{L}\sum_{k=1}^{L-1}f\left(\frac{\pi k}{L}\right)=\frac{1}{\pi}\int_0^\pi dx\, f(x)-\frac{1}{2L}\left[f(0)+f(\pi)\right]+\frac{\pi}{12L^2}\left[f'(\pi)-f'(0)\right]+\mathcal{O}(L^{-4}),
\eal
where $f(x)$ is assumed to be a continuous function in the interval $0\leq x\leq \pi$.
Let us consider a function $F(x_1,x_2)$ well defined at $x_1=x_2=0$ and with the symmetry properties
$F(x_1,x_2)=F(x_2,x_1)=F(\pi-x_1, x_2)$.
Then, double application of the Euler--Maclaurin formula yields
%\begin{widetext}
\bal
\label{suma}
\frac{1}{L^2}\sum_{k_1=1}^{L-1}\sum_{k_2=0}^LF\left(\frac{\pi k_1}{L},\frac{\pi k_2}{L}\right)=&\frac{1}{\pi^2}\int_0^\pi dx_1\int_0^\pi dx_2\,F(x_1,x_2)-\frac{1}{3L^2}\int_0^\pi dx_2\,F_{x_1}(0,x_2)\nn
&-\frac{1}{L^2}F(0,0)-
\frac{\pi}{6L^3}\lim_{x_1\to 0}\left[F_{x_1}(x_1,0)-F_{x_2}(x_1,0)\right]+
\mathcal{O}(L^{-4}),
\eal
%\end{widetext}
where $F_{x_1}(x_1,x_2)\equiv \partial_{x_1} F(x_1,x_2)$ and $F_{x_2}(x_1,x_2)\equiv \partial_{x_2} F(x_1,x_2)$. Because of the symmetry properties of $F$, one has $F_{x_2}(x_1,x_2)=F_{x_1}(x_2,x_1)$. Therefore, if the value of $F_{x_1}(x_1,x_2)$ at $x_1=x_2=0$ is well defined [i.e., $\lim_{x_1\to 0}\lim_{x_2\to 0}F_{x_1}(x_1,x_2)=\lim_{x_2\to 0}\lim_{x_1\to 0}F_{x_1}(x_1,x_2)$], then  the term of $\mathcal{O}(L^{-3})$ in Eq.\ \eqref{suma} vanishes.
On the other hand, if  $\lim_{x_1\to 0}\lim_{x_2\to 0}F_{x_1}(x_1,x_2)\neq  \lim_{x_2\to 0}\lim_{x_1\to 0}F_{x_1}(x_1,x_2)$, the term of $\mathcal{O}(L^{-3})$ is  not necessarily zero.

\subsubsection{$T_{3A,1}$}
We now identify $F(x_1,x_2)$ with the function defined by Eq.\ \eqref{C23} and apply the Euler--Maclaurin formula \eqref{suma} to $T_{3A,1}$ [cf.\ Eq.\ \eqref{C24}]. For details see e.g. \cite{WW96}. The result is
\beq
\label{T3A1}
T_{3A,1}=t_{3A,1}^{(0)}+\frac{t_{3A,1}^{(2)}}{L^{2}}+\frac{t_{3A,1}^{(3)}}{L^3}+\mathcal{O}(L^{-4}),
\eeq
with
\begin{subequations}
\bal
t_{3A,1}^{(0)}=&\frac{3}{2\pi^2}\int_0^{\pi}dx_2\int_0^{\pi}dx_1  F\left(x_1,x_2\right)\nn
=&-1.85021305488\cdots,
\eal
\beq
t_{3A,1}^{(2)}=\frac{12+5\sqrt{2}}{4\pi}.
\eeq
\end{subequations}
Since $F_{x_1}(x_1,x_2)$ is not univocally defined at $x_1=x_2=0$ [in fact,  $\lim_{x_1\to 0}\lim_{x_2\to 0}F_{x_1}(x_1,x_2)-  \lim_{x_2\to 0}\lim_{x_1\to 0}F_{x_1}(x_1,x_2)=-\frac{1}{3}$], it turns out that the coefficient $t_{3A,1}^{(3)}$ is different from zero. An {empirical} fit gives
\beq
\label{t_3A,1(3)}
t_{3A,1}^{(3)}=0.6181(1).
\eeq
The digit enclosed by parentheses is a conservative estimate of the error bar in the last digit. The fit \eqref{t_3A,1(3)} is obtained by evaluating $\left[T_{3A,1}-t_{3A,1}^{(0)}-{t_{3A,1}^{(2)}}{L^{-2}}\right]L^3$ at $L=500$, $1000$, $2000$, and $3000$, which yields $0.618054$, $0.618059$, $0.618073$, and $0.618085$, respectively. A quadratic extrapolation to $L\to\infty$ gives the value $0.618091$. The same procedure confirms that the remainder in Eq.\ \eqref{T3A1} is indeed $\mathcal{O}(L^{-4})$.
\subsubsection{$T_{3A,2}$}
Next, we consider the term $T_{3A,2}$ defined by Eq.\ \eqref{C25}. We eliminate one of the summations by means of the Euler--MacLaurin formula \eqref{EM}. The result is
\beq
\label{T3A2}
T_{3A,2}
=S_1 +S_2+\frac{\epsilon}{L}+\mathcal{O}(L^{-4}),
\eeq
where
\begin{subequations}
\beq
S_1\equiv \frac{3}{\pi L}\sum_{k =1}^{L-1}\left(\frac{1}{k}-2\ln\frac{\pi k }{L}\right),
\eeq
\bal
S_2\equiv\frac{6}{\pi L}\sum_{k =1}^{L-1}\ln\left[\pi+\sqrt{\pi^2+\left(\frac{\pi k }{L}\right)^2}\right]+\frac{3}{L^2}\sum_{k =1}^{L-1}\frac{1}{\sqrt{\pi^2+(\frac{\pi k }{L})^2}}
-\frac{\pi^2}{2L^3}\sum_{k =1}^{L-1}\frac{1}{\left[\pi^2+(\frac{\pi k }{L})^2\right]^{3/2}},
\eal
\end{subequations}
and $\epsilon/L$ accounts for small contributions to $T_{3A,2}$ not captured by the Euler--Maclaurin formula. The numerical value of $\epsilon$ can be {empirically} measured as
\beq
\label{eps}
\epsilon=0.003510438(1).
\eeq
This is obtained by evaluating  $\left[T_{3A,2}-S_1 -S_2\right]L$ up to $L=500$ and extrapolating to $L\to\infty$. This also confirms that the remainder in Eq.\ \eqref{T3A2} is $\mathcal{O}(L^{-4})$.

The sum $S_1$ can be explicitly evaluated as
\bal
\label{S1}
S_1=&\frac{3}{\pi L}\Bigg[\ln L+\gamma_E-\frac{1}{2L}-\frac{1}{12L^2}+\mathcal{O}(L^{-4})+2(L-1)\ln\frac{L}{\pi}-2\ln(L-1)!\Bigg]\nn
=&\frac{3}{\pi}\Bigg[2-\ln\pi^2+\frac{\gamma_E+\ln\frac{\pi}{2}}{L}-\frac{2}{3L^2}-\frac{1}{12L^3}+\frac{1}{180L^4}\Bigg]+\mathcal{O}(L^{-5}),
\eal
where $\gamma_E$ is the Euler constant.

As for $S_2$, reapplication of the Euler-Maclaurin formula yields
\bal
\label{S2}
S_2&=\frac{6}{\pi^2}\int_0^\pi dx \ln\left[\pi+\sqrt{\pi^2+x^2}\right]-\frac{3}{\pi L}\ln[2(1+\sqrt{2})\pi^2]+\frac{1}{2\pi (2+\sqrt{2})L^2}
+\frac{3}{\pi L}\int_0^\pi \frac{dx}{\sqrt{\pi^2+x^2}}\nn
&
-\frac{3(1+\sqrt{2})}{2\sqrt{2}\pi L^2}-\frac{1}{8\sqrt{2}\pi L^3}-\frac{\pi}{2L^2}\int_0^\pi \frac{dx}{\left(\pi^2+x^2\right)^{3/2}}+\frac{1+2\sqrt{2}}{8\sqrt{2}\pi L^3}+\mathcal{O}(L^{-4})\nn
=&\frac{6}{\pi}\left\{\ln\left[(1+\sqrt{2})^2\pi\right]-1\right\}-\frac{3\ln 2\pi^2}{\pi L}-\frac{4+5\sqrt{2}}{4\pi L^2}+\frac{1}{4\pi L^3}+
\mathcal{O}(L^{-4}).
\eal

Insertion of Eqs.\ \eqref{S1} and \eqref{S2} into Eq.\ \eqref{T3A2} gives
\beq
\label{T3A2f}
T_{3A,2}=t_{3A,2}^{(0)}+\frac{t_{3A,2}^{(1)}}{L}+\frac{t_{3A,2}^{(2)}}{L^2}+\mathcal{O}(L^{-4}),
\eeq
with
\begin{subequations}
\beq
t_{3A,2}^{(0)}=\frac{12}{\pi}\ln\left(1+\sqrt{2}\right)=3.366599\cdots,
\eeq
\beq
t_{3A,2}^{(1)}=-\frac{3}{\pi}\left(\ln 4\pi-\gamma_E\right)+\epsilon= -1.862239324(1)
\eeq
\beq
t_{3A,2}^{(2)}=-\frac{12+5\sqrt{2}}{4\pi}=-t_{3A,1}^{(2)}.
\eeq
\end{subequations}

\subsubsection{$T_{3A}$}
Combining Eqs.\ \eqref{T3A1} and \eqref{T3A2f} in Eq.\ \eqref{T3A}, one finds
\beq
\label{TT3A}
T_{3A}=t_{3A,1}^{(0)}+t_{3A,2}^{(0)}+\frac{t_{3A,2}^{(1)}}{L}+\frac{t_{3A,1}^{(3)}}{L^3}+\mathcal{O}(L^{-4}).
\eeq

\subsection{Final results}

From Eqs.\ \eqref{Q1}, \eqref{Q3(1)}, \eqref{RR2B}, \eqref{RR3B}, and \eqref{TT3A}, we finally obtain Eq.\ \eqref{Q31body}, where
\begin{subequations}
\beq
b_0=t_{3A,1}^{(0)}+t_{3A,2}^{(0)}=1.5163860\cdots,
\eeq
\beq
b_1=\frac{1}{2}+3r_{2B}^{(0)}+r_{3B}^{(1)}+t_{3A,2}^{(1)}= -1.354709757(1),
\eeq
\beq
b_3=-\frac{1}{2}+3r_{2B}^{(2)}+r_{3B}^{(3)}+t_{3A,1}^{(3)}=0.2574(1).
\eeq
\end{subequations}
Note that a rather compact expression for $b_0$ is
\bal
b_0=&\frac{3}{2\pi^2}\int_0^\pi dx_2\int_0^\pi dx_1 F_0(x_1,x_2)\nn
=&\frac{6}{\pi^2}\int_0^{\pi/2}\frac{dx}{1+\sin^2 x}K\left(\frac{1}{(1+\sin^2 x)^2}\right),
\eal
where $K(x)$ is the complete elliptic integral of the first kind [cf. Eq.\ \eqref{elliptic}].
It is worth noticing that the nonempirical contribution to the coefficient $b_1$ (i.e, without $\epsilon$) accounts for $99.74$\% of its numerical value.

\section{Integral representation of $P_d(z)$}
\label{app5}
By using the identity
\beq
\int_0^\infty d\zeta\,\exp\left[-\zeta\left(\frac{d}{z}-\sum_{i=1}^d C_i\right)\right]=\left(\frac{d}{z}-\sum_{i=1}^d C_i\right)^{-1},
\eeq
Eq.\ \eqref{arbdgenfunc} can be rewritten as
\bal
P_d(z)=&\frac{d}{z L^d}\int_0^\infty d\zeta\,e^{-\zeta d/z}\sum_{k_1=0}^{L-1}e^{\zeta C_1}\sum_{k_2=0}^{L-1}e^{\zeta C_2}\cdots\sum_{k_d=0}^{L-1}e^{\zeta C_d}\nn
=&\frac{d}{z}\int_0^\infty d\zeta\,e^{-\zeta d/z}\left(\frac{1}{L}\sum_{k=0}^{L-1}e^{\zeta \cos(2\pi k/L)}\right)^d.
\eal
The last term is precisely the representation \eqref{intrepP}.

It is instructive to rederive Eq.\ \eqref{intrepP} in a different way, which illustrates how results for a periodic lattice can be obtained from known results for the infinite lattice. The starting point is the generating function $\PP^{(\infty)}(\overrightarrow{\ell }, z)$ for the sojourn probability at site $\overrightarrow{\ell }=(\ell _{1} ,\ldots,\ell _{d} )$ on an \textit{infinite} hypercubic lattice. This lattice Green function is known to have the exact integral form \cite[p.\ 146]{H95}
\beq
\label{infgenfun}
\PP^{(\infty)}(\overrightarrow{\ell },z)=\frac{d}{z} \int _{0}^{\infty }d\zeta\,e^{-\zeta d/z}  \prod _{j=1}^{d}I_{\ell_{j} }  (\zeta ) ,
\eeq
where $I_{n} (\cdot )$  stands for the $n$th-order modified Bessel function of the second kind. Clearly, a hypertorus of linear size $L$ can be thought of as an infinite lattice that has been wrapped onto itself an infinite number of times. Thus, the probability of return to the origin of a P\'olya walker on a hypertorus can be computed as a infinite sum of contributions stemming from walker trajectories on an \textit{infinite} lattice, whereby the walker advances a multiple of the linear lattice size $L$ along every spatial direction.  In other words,
\bal
P_d(z)\equiv \PP(\overrightarrow{0},z)=&\sum _{\overrightarrow{\ell }}\PP^{(\infty)}(\overrightarrow{\ell }L,z)\nn
 =& \sum _{\ell _{1}=-\infty }^{\infty }\sum _{\ell _{2} =-\infty }^{\infty }\cdots \sum _{\ell _{d-1} =-\infty }^{\infty }\sum _{\ell _{d} =-\infty }^{\infty } \PP^{(\infty)} (\ell _{1} L,\ell _{2} L,\ldots,\ell _{d-1} L,\ell _{d} L,z).
\eal
(In this case, the sum includes the origin $\vec{\ell }=0$). Now, using the well known identity \cite[p.\ 695 ]{PBM88}
\beq
\sum _{k=0}^{\infty }I_{kL} (z) =\frac{I_{0} (z)}{2} +\frac{1}{2L} \sum _{k=0}^{L-1}e^{z\cos (2\pi k/L)}
\eeq 	
in Eq.\ \eqref{infgenfun}, and switching to our short hand notation, we arrive at the result  \eqref{intrepP}. Making the replacement $z=1-s$ in \eqref{intrepP} and inserting the result into Eq.\ \eqref{genres}, one finds a general expression for the walklength of a mortal walker in the presence of a deep trap and background trapping. The above integral does not seem expressible in a closed form except in the case $d=1$, leading to the well-known result [cf.\ Eq.\ \eqref{Q1z}]:
\beq
P_1(z)=\frac{1}{\sqrt{1-z^{2} } } \frac{1+\left[X(z)\right]^{N} }{1-\left[X(z)\right]^{N}}.
\eeq

\end{document}